\documentclass[superscriptaddress,prl,floats,amsmath,amssymb,aps,twocolumn,showpacs]{revtex4}
\usepackage{graphicx}
\usepackage{amsmath}
\usepackage{amssymb}
\usepackage{dcolumn}
\usepackage{color}
\begin{document}
\title{Hydrogenation Facilitates Proton Transfer Through Two-Dimensional Honeycomb Crystals}
\author{Yexin Feng}
\thanks{These two authors contributed equally}
\affiliation{School of Physics and Electronics, Hunan University, Changsha 410082, P. R. China}
\author{Ji Chen}
\thanks{These two authors contributed equally}
\affiliation{Thomas Young Centre, London Centre for Nanotechnology, and Department of Physics and Astronomy, University
College London, London WC1E 6BT, United Kingdom}
\author{Wei Fang}
\affiliation{Thomas Young Centre, London Centre for Nanotechnology, and Department of Chemistry, University
College London, London WC1E 6BT, United Kingdom}
\author{En-Ge Wang}
\affiliation{School of Physics, ICQM, and Collaborative Innovation Center of Quantum Matter,
Peking University, Beijing 100871, P. R. China}
\author{Angelos Michaelides}
\email{angelos.michaelides@ucl.ac.uk}
\affiliation{Thomas Young Centre, London Centre for Nanotechnology, and Department of Physics and Astronomy, University
College London, London WC1E 6BT, United Kingdom}
\author{Xin-Zheng~Li}
\email{xzli@pku.edu.cn}
\affiliation{School of Physics, ICQM, and Collaborative Innovation Center of Quantum Matter,
Peking University, Beijing 100871, P. R. China}
\date{\today}

\begin{abstract}
%
%Although calculations have shown pristine graphene is impermeable to proton at room temperature,
Recent experiments have triggered a debate about the ability of protons to transfer through
individual layers of graphene and hexagonal boron nitride (h-BN).
However, calculations have shown that the barriers to proton penetration can, at more than 3 eV, be excessively high.
Here, on the basis of first principles calculations, we show that the barrier for proton penetration is
significantly reduced, to less than 1 eV, upon hydrogenation even in the absence of pinholes in the
lattice.
Analysis reveals that the barrier is reduced because hydrogenation destabilises the initial state (a
deep-lying chemisorption state) and expands the honeycomb lattice through which the protons penetrate.
This study offers a rationalization of the fast proton transfer observed in
experiments, and highlights the ability of proton transport through single-layer materials in
hydrogen rich solutions.
\end{abstract}

\pacs{68.43.Bc, 82.45.Mp, 71.15.Pd, 82.65.+r}

%68.43.Bc: Ab initio calculations of adsorbate structure and reactions (for electronic  structure  of  adsorbates,  see  73.20.Hb;  for  adsorbate reactions,  see  also  82.65.+r  Surface  and  interface  chemistry; heterogeneous catalysis at surfaces
%
%82.45.Jn: Surface structure, reactivity and catalysis (see also 82.65.+r Surface and interface chemistry; heterogeneous catalysis at surfaces)
%
%71.15.Pd: Molecular dynamics calculations (Car-Parrinello) and other numerical simulations
%
%82.65.+r: Surface and interface chemistry; heterogeneous catalysis at surfaces (for temporal and spatial patterns in surface reactions, see 82.40.Np; see also 82.45.Jn Surface structure, reactivity and catalysis in electrochemistry; see also 68.43.-h Chemisorption/physisorption: adsorbates on surfaces)

\maketitle

\clearpage

Selective sieving of ions and molecules through thin membranes is a key step for a
wide range of applications such as water purification and ion exchange membrane fuel cells{~\cite{Geim2014,Geim2016,Graphene-oxide-Manchester,Surwade2015,Cohen-Tanugi2012,Zilman2009,OHern2012,Celebi2014,Achtyl2015,OHern2016,Joshi2014,Mi2014,Kim2013science,ZengXC2015,Cheng2016SciAdv,Bocquet2016}.
Two-dimensional (2D) materials like graphene and hexagonal boron nitride (h-BN) offer potential as membrane materials since they are
 a single atom thick and have high mechanical stability and  flexibility{~\cite{Geim2014,Geim2016,Surwade2015,Cohen-Tanugi2012,OHern2012,Celebi2014,Achtyl2015,OHern2016}}.
For some time, it was believed that pristine graphene and h-BN were impermeable to ions
due to high energy barriers for penetration~\cite{Miao2013,NJP2010}.
Recent experiments, however, have suggested that protons can in fact penetrate
 pristine graphene and h-BN~\cite{Geim2014,Geim2016}.
% a suggestion that has led to intense
%discussions of the underlying mechanism~\cite{Geim2014,Poltavsky2016,Isotope-2016-jpcl,H-BN-2017-pccp}.
%
In the measurements, the 2D materials were immersed in proton conducting polymers or aqueous solutions, and from temperature ($T$)-dependent proton conductivity measurements, proton penetration barriers of only 0.8 and 0.3 eV were estimated for single-layer graphene and h-BN, respectively~\cite{Geim2014}.
Note that these estimated barriers include contributions from zero-point energy (ZPE)~\cite{Geim2016}.
Defects such as atomic pinholes are known to facilitate proton transfer~\cite{Achtyl2015}.
A certain level of defects will inevitably be present, associated e.g. with sp$^3$ carbon
atoms~\cite{nanolett-manchester}.
However in Refs.~\cite{Geim2014,Geim2016}, various measurement techniques
(transmission/tunnelling electron microscopy, Raman spectroscopy and measurements of gas leakage) were used to
support the assertion that the proton transfer mechanism was not facilitated by atomic defects in
the membranes.
Considerable theoretical effort has been devoted towards understanding the microscopic details of
how protons penetrate 2D materials~\cite{Miao2013,NJP2010,Walker2015,Achtyl2015,Isotope-2016-jpcl,H-BN-2017-pccp}.
It has been established on the basis of density-functional theory (DFT) calculations that the barriers to proton
penetration through pristine graphene and h-BN in vacuum can be excessively high.
Specifically, computed barriers of 3.5-4.0 eV have been reported for chemisorbed protons (i.e.\ protons that
are covalently bonded to the 2D materials) to penetrate graphene~\cite{NJP2010,Miao2013,Poltavsky2016}.
If the protons do not chemisorb on the surface but rather penetrate the sheet via a metastable physisorption state,
smaller barriers of 1.4-2.6 eV have been reported~\cite{NJP2010,Miao2013,Poltavsky2016}.
However, the physisorption state is only a very shallow minimum, separated from the much more stable chemisorption
state by a barrier of $\le$ 0.1 eV~\cite{Davidson2014}.
Therefore, it seems unlikely that penetration from the physisorption state is the dominant mechanism for
fast proton conduction~\cite{Davidson2014,H-BN-2017-pccp}.
Nonetheless this indicates that hydrogenation of graphene is facile and that graphene sheets immersed in proton conducting polymers or aqueous solutions could be hydrogenated or protonated to some extent.
In addition, given the light mass of the proton, the role of nuclear quantum effects (NQEs) such as
tunnelling and zero point motion could be relevant to the process, as shown e.g. through two interesting
recent computational studies~\cite{Poltavsky2016, Isotope-2016-jpcl}.
In this Letter, we report a study of proton transfer through graphene and h-BN, focusing on
the transmission mechanism.
Consistent with earlier studies, a very high potential energy barrier of $\sim$3.6 eV is found for proton penetration of graphene via the chemisorption state.
Using \textit{ab initio} path-integral molecular dynamics (PIMD)~\cite{PIMD1,PIMD2,PIMD3,Guo2016,Tuckerman2001,ZhangQ2008,Walker2010}, we take into account nuclear quantum effects (NQEs) and finite temperature thermal effects.
We find that NQEs reduce the penetration barrier of graphene by 0.46 eV ($12\%$) at 300~K,
which is unlikely to be responsible for the experimentally observed high transfer rate.
Upon considering the role $\text{sp}^3$ bonded atoms play on the penetration process, created here
by hydrogenation of graphene and h-BN, we find that hydrogenation can reduce the penetration
barriers significantly to less than 1 eV.
This reduction arises because the hydrogenation induced $\text{sp}^2$ to $\text{sp}^3$ transformation
destabilises the deep-lying chemisorption state in which the proton can get trapped on the pristine
membranes.
Geometrically, hydrogenation also expands the six-atom rings through which protons transfer.
Analysis of the penetration barriers associated with many distinct hydrogenated membranes reveals a clear correlation between the height of the penetration barriers and the
local degree of hydrogenation at the proton transfer site.
Overall this work highlights the significant difference in proton penetration barriers that can be found
in the vicinities of $\text{sp}^3$ bonded atoms and helps to rationalise the facile transport of
protons through single-layer materials.
Our DFT calculations were performed using the Vienna ab initio Simulation Package (VASP)~\cite{vasp1996}, with an in house implementation
of the \textit{ab initio} constrained-centroid MD/PIMD methods~\cite{Chen2013NC,Guo2016}.
The optB88-vdW functional was chosen in the electronic structure calculations so as to obtain a
good description of the hydrogen (H)-bonding interactions and dispersion
forces~\cite{Klimes2010,Klimes2011}.
Charged cells were employed to describe the protons in the simulations, and we confirmed that any charge states considered were correctly characterized with Bader analysis~\cite{bader-charge-book,bader-charge}.
We hydrogenated graphene to varying degrees without generating pinholes, using supercells ranging
from 4$\times$4 to 8$\times$8.
After hydrogenation, the supercell shape and size was allowed to change.
For each partially hydrogenated structure, we have considered the two lowest energy structures following the study of
Ref.~\cite{H_cluster_prb}.
The climbing image nudged elastic band (cNEB) method was used in calculating the static penetration
barriers~\cite{cNEB2000}, with a force convergence criterion of 0.03 eV $\text{\AA}^{-1}$ and
all atoms were allowed to relaxed.
Beyond the static description, the classical and quantum free energy profiles were obtained with
constrained MD and PIMD approaches~\cite{Tuckerman2001,ZhangQ2008,Walker2010}; with the
constraint applied on the vertical distance of the proton from the 2D layer.
%\textcolor{red}{The proton was constrained at different vertical positions from the 2D layer.}
%
%$T$ was targeted at 300~K by the Anderson thermostat.
%
A 0.5 fs time step was used and the imaginary-time path in the PIMD simulations was sampled with 48 replicas, at a target temperature of 300~K.
After thermalization, 30,000 steps (15 ps) were collected to calculate the constraint force, for each constraint point.
By integrating over the constraint forces, the free energy profiles were obtained as detailed in the supplementary information (SI).
On free-standing graphene protons adsorb preferentially at the chemisorption site directly above a carbon atom (Fig.~\ref{fig1} (a)).
% the chemisorption site is more stable (Fig.~\ref{fig1} (a)).
%
%The proton can easily transfer from the physisorption site to the more stable chemisorption one
%with a barrier of $\le$ 0.1 eV~\cite{Davidson2014}.
%
From the chemisorption site, our calculations yield a proton penetration barrier of
3.60 eV.
As noted, in previous experiments, the 2D layers were surrounded by proton conducting polymers or aqueous solutions~\cite{Geim2014,Geim2016,Walker2015,Achtyl2015,H-BN-2017-pccp}.
In the current study, we do not aim to model such an aqueous environment.
However, in order to gain an initial understanding of how the presence of water might impact upon the proton penetration process,
we employed the simplified model shown in the
inset of Fig.~\ref{fig1} (b).
This model contains one water molecule on each side of the graphene layer and
with the addition of a proton it enables us to model proton transfer from an $\text{H}_\text{3}\text{O}^+$ on one
side of the sheet to an $\text{H}_\text{2}\text{O}$ on the other side of the sheet~\cite{Kurita2008}.
Our calculations show that the proton adsorbs at either the water molecule or the chemisorption
site of the graphene sheet with very similar stability (Fig.~\ref{fig1} (b)).
%
%As shown in Fig. 1 (b), the proton adsorbs at either the water molecule or the chemisorption site of the graphene sheet with very close stability.
%
The metastable physisorption site for protons on free-standing graphene, as illustrated in
Fig.~\ref{fig1} (a), disappears due to the presence of water.
The energy barrier for a proton to transfer from the $\text{H}_\text{3}\text{O}^+$ to the chemisorption
site is less than 0.1 eV.
The penetration barrier from the chemisorption site is 3.65 eV when water is present, very similar to the 3.60 eV obtained in the absence of water.
The energy differences between physisorbed water molecules on different sites or with different orientations
are only a few meV~\cite{Kurita2008,MaJie2011,Hamada2012}, so different configurations of
water molecules will not obviously influence the energy profile of the proton penetration process.
Therefore, in agreement with recent work~\cite{H-BN-2017-pccp}, we conclude that the presence of
water molecules is unlikely to change the fact that very high barriers exist for
protons to transfer across the graphene layer.

\begin{figure}[ht]
\includegraphics[width=1.0\linewidth]{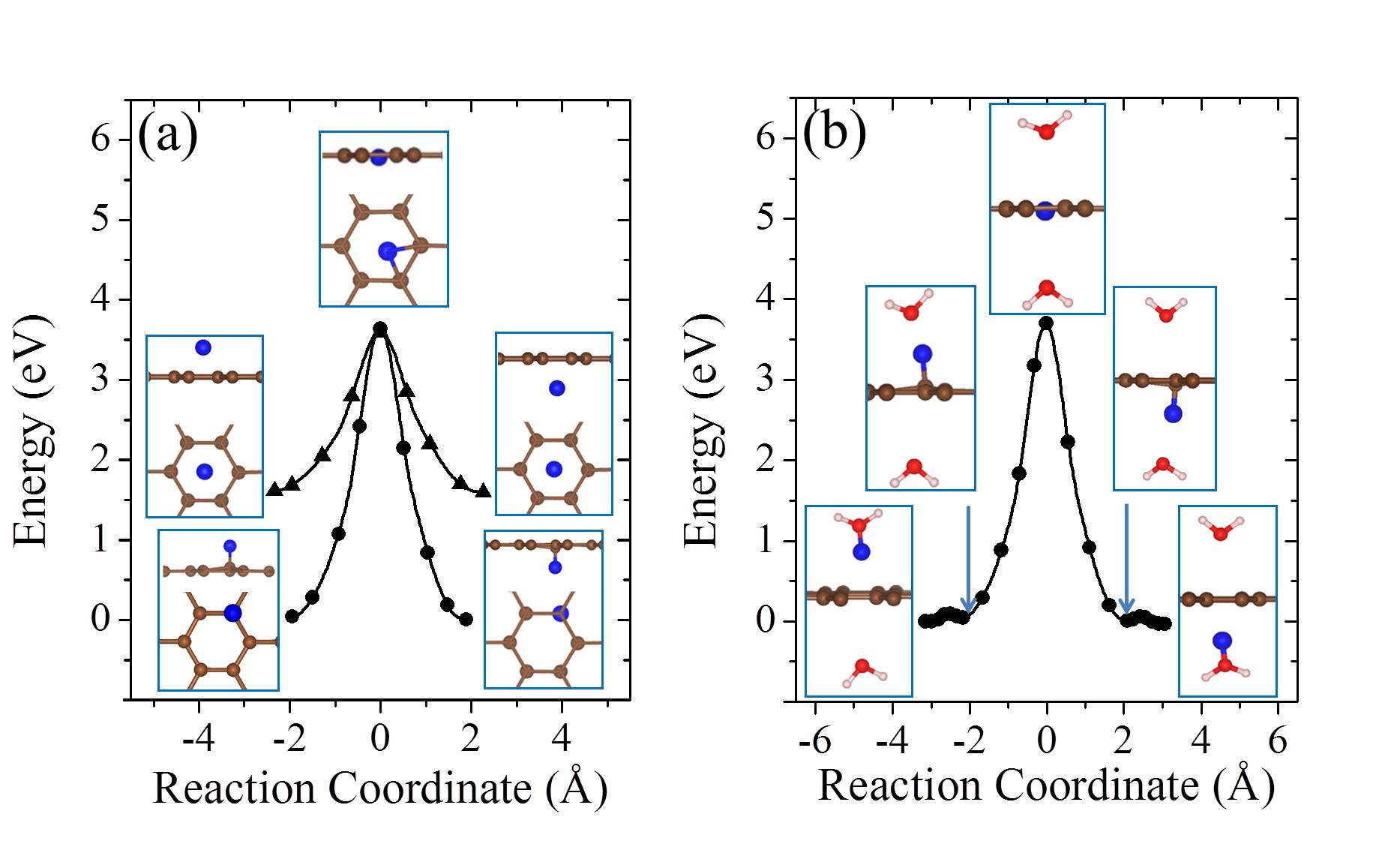}
\caption{\label{fig1}
High barriers for proton transfer through pristine graphene.
Calculated energy profiles as obtained from cNEB calculations, for proton transfer across (a) pristine graphene, and (b) graphene with adsorbed water molecules.
Two energy profiles are shown in (a), one between the metastable physisorption states (upper curve) and one between the chemisorption states (lower curve).
The insets show some of the key states involved in the proton transfer processes.
Brown (red, pink) balls are C (O, H) atoms.
Protons are represented by blue balls.
}
\end{figure}
Finite temperature and NQEs [ZPE and quantum tunneling] are known to alter the barriers of chemical processes, particularly proton transfer barriers.
To understand the importance of such effects on the current system, we performed a series of \textit{ab initio} MD and PIMD simulations from which free energy barriers for proton penetration were obtained.
The results of these simulations are shown in Fig.~\ref{fig2}.
We find that the pure thermal effects on the barrier are relatively small and the free energy
barrier in the model containing a water on either side of the sheet is 3.70 eV at 300 K.
When NQEs are accounted for with PIMD the barrier is reduced by 0.46 eV at 300 K (Fig.~\ref{fig2}), quite a substantial reduction.
Analysis reveals that this reduction in the free energy barrier is due to enhanced quantum delocalisation
of the proton at the transition state compared to the initial state.
This is similar behavior to that observed for H chemisorption on graphene~\cite{Davidson2014},
and is illustrated by the snapshots shown in Fig.~\ref{fig2}.
The reduction arising from NQEs is also in line with that reported by Poltavsky \textit{et al.} when
similar PIMD methods are used~\cite{Poltavsky2016}, although a different computational model and reaction
pathway was considered by Poltavsky \textit{et al.}.
However, considering the fact that the free energy barrier for the process examined remains $>$ 3 eV at 300 K, we
conclude that NQEs alone can't rationalise the experimentally observed fast proton transfer.
%
%We note that we also examined how isotope effects impact upon the penetration process and
%show a close-up of the free-energy profiles in Fig. ~\ref{fig2} (b).
%
%For the specific penetration process considered, upon replacing D by H, the
%penetration barrier reduces by 134 meV.
%
%In Ref.~\cite{Geim2016}, the experimentally observed difference of activation barrier for proton and deuteron was reported to be about 60 meV.
%

\begin{figure}[ht]
\includegraphics[width=0.8\linewidth]{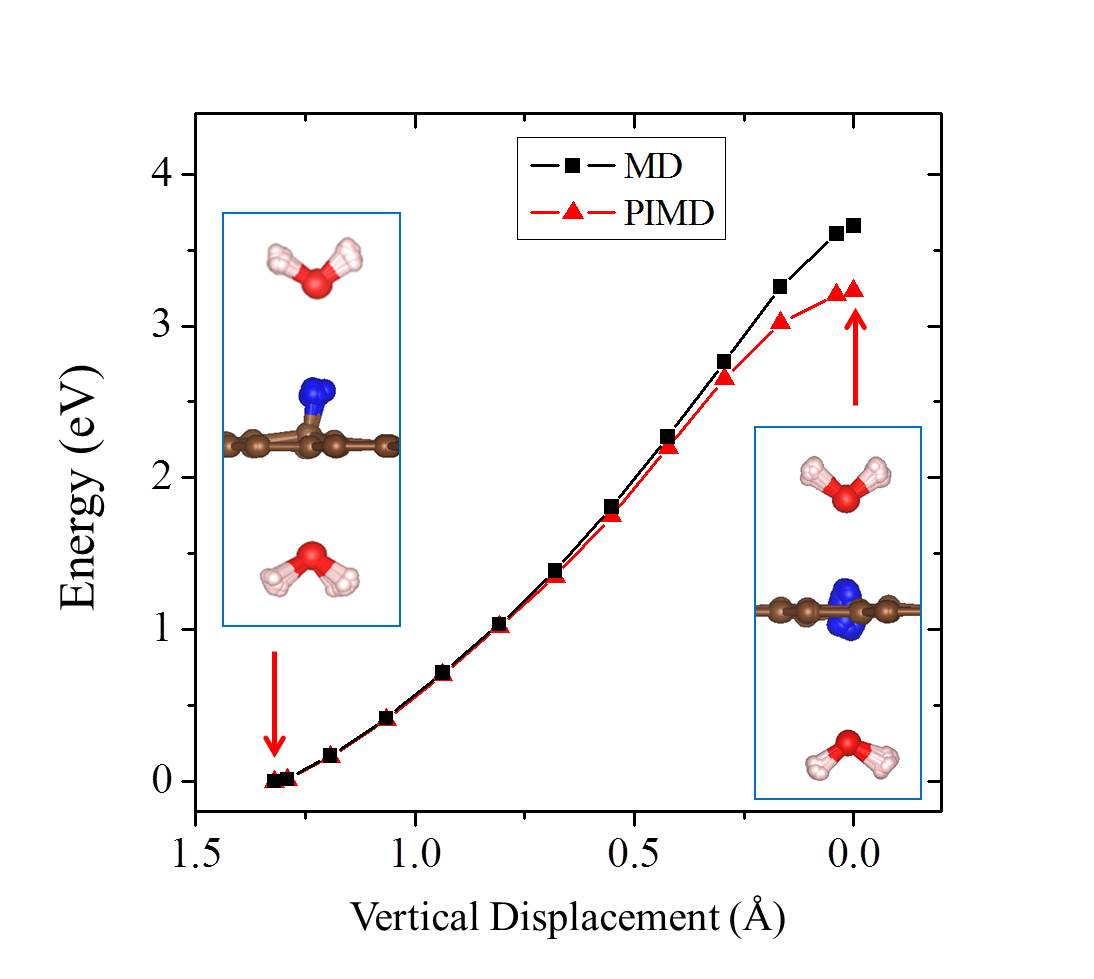}
\caption{\label{fig2}
Free energy profiles at 300 K obtained with \textit{ab initio} constrained MD and PIMD simulations for proton transfer across a graphene sheet in the presence of water molecules.
The MD simulations take into account thermal effects, whereas the PIMD simulations capture thermal and nuclear quantum effects.
PIMD simulation snapshots for the initial state and transition state are also shown.
Blue (red and pink) balls represent the beads of protons (O and H atoms), for one snapshot in a PIMD simulation.
The centroids of the C atoms are shown as brown balls.
%Corresponding constraint forces from which the free-energy profile are calculated. Inset is the atomic structure of simulation model used in cNEB and constraint MD/PIMD calculations.
%
%%With the six C atoms surrounding the transfer channel hydrogenated, cNEB calculation results of proton transfer is illustrated along with atomic structures of initial, final and transition state. Brown (red, pink) balls are C (O, H) atoms.
%
%%Blue balls are Proton nuclei.
%
}
\end{figure}
We noted in the introduction that carbon atoms with $\text{sp}^3$ character are invariably present even
in pristine graphene~\cite{nanolett-manchester}.
With this in mind we explored how the presence of chemisorbed hydrogens impact the proton penetration
barrier of graphene.
Adsorbed hydrogens are examined since when they chemisorb they lead to an $\text{sp}^2$ to $\text{sp}^3$
hybridisation of the carbon atoms they are bonded to but also because the hydrogenation of graphene is
facile~\cite{hydrogenation-graphene}.
A broad range of hydrogenation scenarios was considered ranging from having just a single chemisorbed
hydrogen at a proton penetration site to fully hydrogenated graphene (graphane) sheets.
Examples of some of the structures considered are shown in Fig.~\ref{fig3}, with full details given in the SI~\cite{SI}.
Upon computing the proton penetration barriers through the various hydrogenated and partially hydrogenated
sheets considered, we find that hydrogenation leads to reduced proton penetration barriers. The actual barriers obtained depends sensitively on the particular hydrogenation structure, with barriers for some hydrogenated structures reduced very substantially to $<$ 1 eV.

\begin{figure}[ht]
\includegraphics[width=0.85\linewidth]{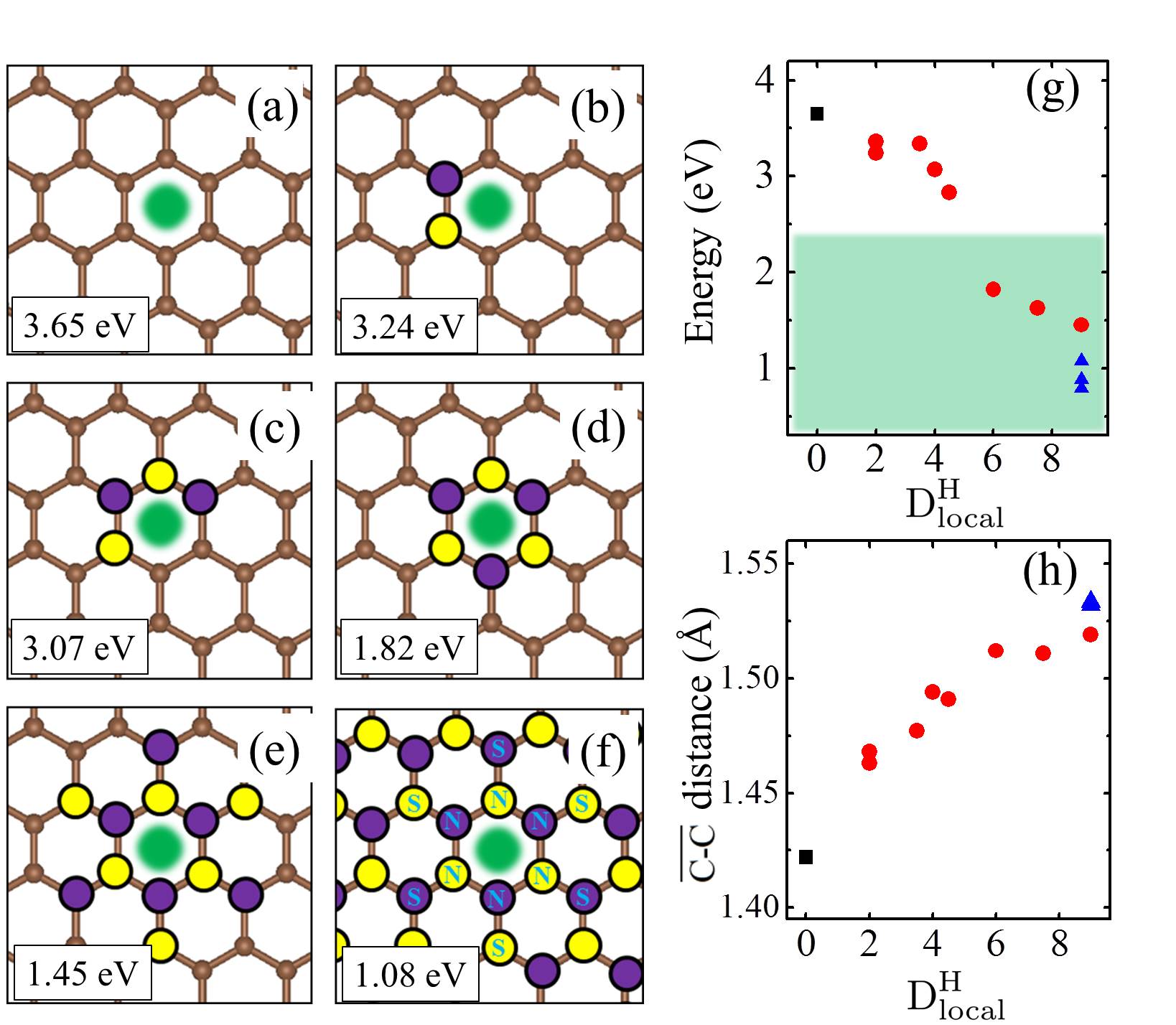}
\caption{\label{fig3}
Hydrogenation facilitates proton penetration through graphene.
(a)-(f) The atomic structures of graphene and a selection of hydrogenated graphene sheets with
different degrees of local hydrogenation.
cNEB energy barriers for proton transfer through each sheet are also reported.
Yellow and violet dots indicate C atoms hydrogenated from the top and bottom sides, respectively.
The large smeared green ball indicates the hole of the C ring through which the proton penetration.
In (f) the nearest neighbor (N) and second nearest neighbor (S) C atoms to the penetration site are indicated.
(g) cNEB barriers as a function of the degree of local hydrogenation ($\text{D}^\text{H}_\text{local}$), for various hydrogenated structures(see Figs. S4 and S5 for details of the structures). The green shaded area indicates low penetration barriers with high $\text{D}^\text{H}_\text{local}$.
(h) The averaged C-C bond distance of the six-C ring through which the proton penetrates as a function of $\text{D}^\text{H}_\text{local}$.
In (g) and (h), the black square, red dots and blue triangles represent data for proton transfer across pristine, partially
hydrogenated and fully hydrogenated graphene, respectively.
}
\end{figure}
Let us now look more closely at the hydrogenated systems and try to understand the barriers obtained.
Crucially we find that the penetration of protons through a single atom layer is a local process and that the height of the penetration barrier depends
primarily on the local degree of hydrogenation in the vicinity of the penetration site.
To show this more clearly, we introduce an order parameter,
$\text{D}^\text{H}_\text{local}$.
$\text{D}^\text{H}_\text{local}$ is defined as $\text{D}^\text{H}_\text{local}=\text{N}^\text{H}_\text{N}+\text{w}\times\text{N}^\text{H}_\text{S}$,
with $\text{N}^\text{H}_\text{N}$ ($\text{N}^\text{H}_\text{S}$) being the number of hydrogenated atoms
at the nearest (second nearest) neighbors of the hole (indicated by N and S in the Fig.~\ref{fig3} (f)), and w representing a weight factor capturing the importance of hydrogenation at the second nearest sites.
The barriers as a function of $\text{D}^\text{H}_\text{local}$, with w set to 0.5, are plotted in Fig.~\ref{fig3} (g) (tuning w from 0.2 to 0.8 gives similar results (Fig. S11)).
Upon computing $\text{D}^\text{H}_\text{local}$ for all barriers considered we found
two interesting features: i) a clear correlation exists between the penetration barrier and
$\text{D}^\text{H}_\text{local}$, with the barrier getting smaller as $\text{D}^\text{H}_\text{local}$ inceases,
and ii) the systems can be categorized into two main groups, with the most significant barrier reduction being found for $\text{D}^\text{H}_\text{local}$ $>$ 6.
Systems with $\text{D}^\text{H}_\text{local}$ smaller than 6 belong to the group with large barriers.
In these systems, the six-C ring through which the proton penetrates is not fully hydrogenated.
C atoms with $\text{sp}^2$ bonding are present in the ring and the proton can chemisorb at these sites before
penetration (Figs. S4 and S5).
It is the presence of the very stable chemisorption sites that lead to particularly high barriers for proton penetration.
For the systems considered, when $\text{D}^\text{H}_\text{local}\ge 6$, the ring is fully hydrogenated.
The $\text{sp}^3$ bonding eliminates the deep-lying chemisorption state before penetration.
In so doing the initial state energy is raised and the barriers are lower than 2.0 eV.
Note that this analysis reveals that because the barrier is related to the local extent of hydrogenation, a sample does not need to have
a very high global degree of hydrogenation for low barrier proton penetration sites to exist.
All that is required is a high local degree of hydrogenation and indeed surface science measurements and previous calculations show that upon hydrogenation there is a tendency for Hs to cluster ~\cite{H_cluster_prl,H_cluster_nanolett,H_cluster_prb}.
Aside from eliminating the chemisorption well, $\text{sp}^3$ bonded carbons also lead to an expansion of the lattice.
This can be seen in Fig.~\ref{fig3} (h) where the averaged C-C distance of a hexagon is shown to increase with $\text{D}^\text{H}_\text{local}$.
This expansion is an additional geometric effect played by $\text{sp}^3$ bonded carbons~\cite{graphane-2010}.
Our simulations with full hydrogenation correspond to the case when the graphene layer is fully
hydrogenated around a local penetration site.
They have the same $\text{D}^\text{H}_\text{local}$ but different barriers in
Fig.~\ref{fig3} (g) (three blue triangles).
To understand why this happens, we take the chair conformation
and a disordered H configuration as examples and show the actual cNEB barrier profiles in Fig.~\ref{fig4}.
The key difference between these two systems is that in the chair conformation, the upper and lower
sides of the graphene sheet are similarly hydrogenated, while in the disordered configuration the two sides of the sheet are hydrogenated to different extents.
Such asymmetric decoration creates structures wherein it is yet more facile for the proton to penetrate form one side to the other.
On a larger scale, one can imagine that hydrogenation can induce different local penetration sites, with the ease of penetration related to the extent of hydrogenation on either side of the sheet.
For h-BN, H atoms also prefer to chemisorb in pairs on B and N atoms, and the averaged binding energy between H atoms and h-BN increases with the degree of hydrogenation~\cite{H-BN-2016-pccp,hBN2000}.
As with graphene, upon examining proton penetration through h-BN we find that the barriers decrease upon
hydrogenation.
As shown in Table I, the barrier through pristine h-BN is as high as 3.33 eV.
In fully hydrogenated h-BN with ordered H configurations (h-BN sheets with stirrup and boat conformations, h-BN$_{\text{stirrup-H}}$ and h-BN$_{\text{boat-H}}$)~\cite{H-BN-apl},
the barrier can be reduced to less than 2.0 eV.
For disordered H configurations, the barrier further decreases to $\sim$0.93 eV.
Some representative energy profiles for partially and fully hydrogenated h-BN sheets are provided in the Figs. S14-S16.
In Table I, we summarize some representative barriers obtained for proton transfer through the various
graphene and h-BN systems considered.
Also included in Table I are the ZPE corrections to the barriers computed within the harmonic approximation.
ZPE effects decrease the barriers to proton penetration in all systems considered and when they are
taken into consideration the lowest barrier on graphene is 0.61 eV and on h-BN it is 0.51 eV.
Finally, we note that we have also considered how substitution of H for D is likely to alter the penetration barriers.
Treating this again at the ZPE level we find a 50 meV difference in penetration barriers between H and D for $\text{G}_\text{disordered-H}$, and a 120 meV difference between H and D for $\text{h-BN}_\text{disordered-H}$.
In each case, the D barrier is slightly larger than the H barrier, in agreement with recent computational work~\cite{Isotope-2016-jpcl} and experiment~\cite{Geim2016}.

\begin{figure}[ht]
\includegraphics[width=1.0\linewidth]{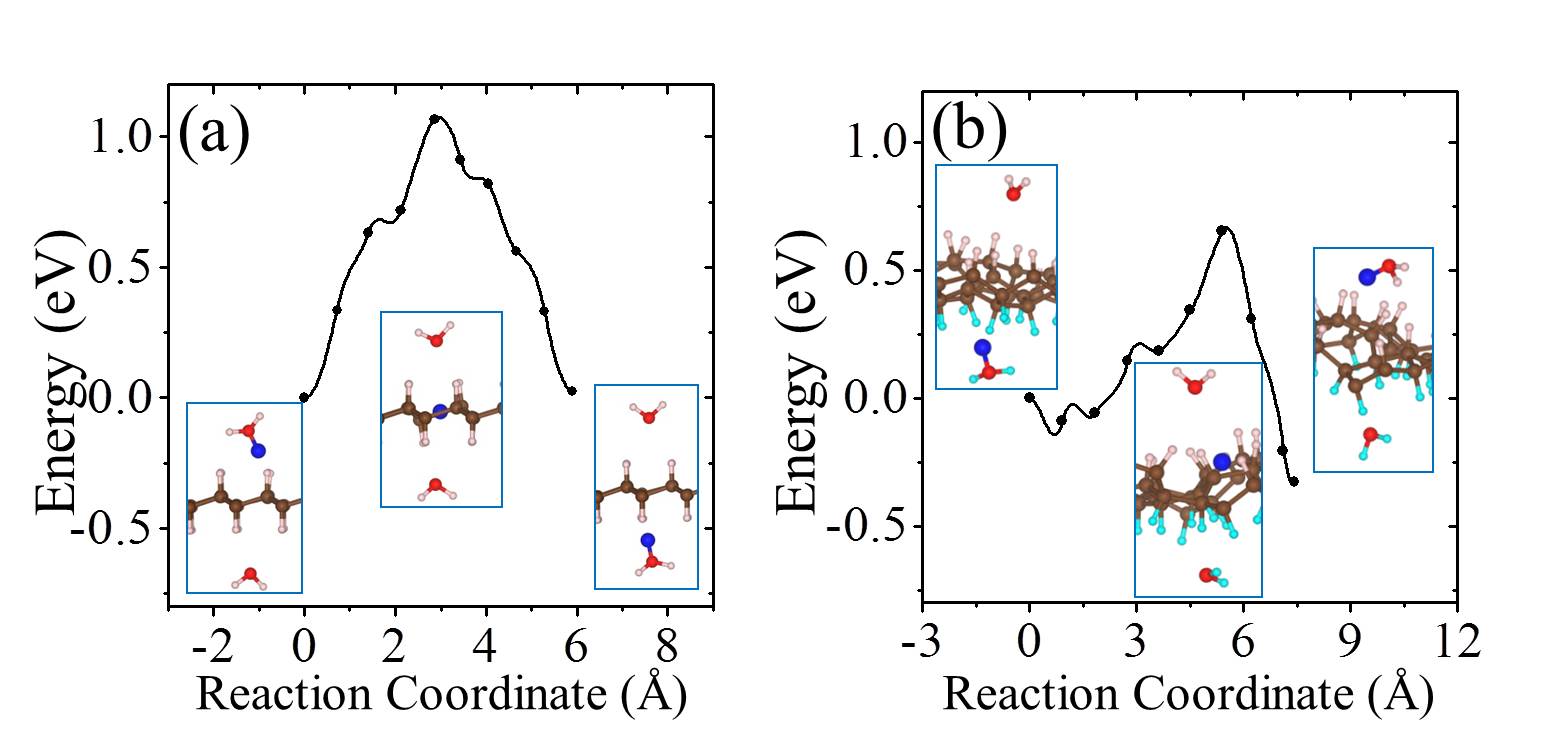}
\caption{\label{fig4}
Energy profiles for proton transfer across fully hydrogenated graphene with (a) the chair conformation and (b) a disordered H configuration.
Insets show the atomic structures for the initial, transition and final states.
Red (pink, brown) balls are O (H, C) atoms.
Protons are represented by blue balls.
For the disordered conformation, the H adatoms below the sheets are colored with cyan (a contrast to pink)
for clarity.
The specific disordered and asymmetric decoration pattern reported in (b) yields a particularly low proton penetration barrier.
Additional information on these structures is given in Fig. S13.
%
%%cNEB results for proton transfer across them are illustrated in (b) and (d), along with atomic structures of initial, final and transition state.
%
}
\end{figure}

\begin{table}[!hbp]
\begin{tabular}{c|c|c|c}
\hline
\hline
      & cNEB Barrier &  ${\bigtriangleup}E_{\text{ZPE}}$ & Barrier \\
\hline
 G$_{\text{pristine}}$ & 3.65 & -0.26 & 3.39  \\
 \hline
 G$_{\text{chair-H}}$ & 1.08 & -0.07 & 1.06  \\
 \hline
 G$_{\text{boat-H}}$ & 0.88 & -0.12 & 0.76  \\
 \hline
 G$_{\text{disordered-H}}$ & 0.79 & -0.18  & \textbf{0.61}  \\
 \hline
 h-BN$_{\text{pristine}}$ & 3.33 & -0.21 & 3.12  \\
 \hline
 h-BN$_{\text{stirrup-H}}$ & 1.43 & -0.53 & 0.90  \\
 \hline
 h-BN$_{\text{boat-H}}$ & 1.91 & -0.35 & 1.56  \\
 \hline
 h-BN$_{\text{disordered-H}}$ & 0.93 & -0.39 & \textbf{0.51}  \\
\hline
\end{tabular}
\caption{Calculated cNEB barrier, ZPE corrections (${\bigtriangleup}E_{\text{ZPE}}$) and corrected barrier (Barrier) for proton transfer across pristine and hydrogenated graphene and h-BN sheets.
The rightmost column should be compared with the experimental values
of 0.8 eV and 0.3 eV in Ref.~\onlinecite{Geim2014}.
${\bigtriangleup}E_{\text{ZPE}}$ is estimated as the ZPE differences between the initial and
transition states. $\text{G}_\text{pristine}$ and $\text{h-BN}_\text{pristine}$ are pristine graphene and h-BN sheets. $\text{G}_\text{chair-H}$, $\text{G}_\text{boat-H}$, $\text{G}_\text{disordered-H}$, $\text{h-BN}_\text{stirrup-H}$, $\text{h-BN}_\text{boat-H}$ and $\text{h-BN}_\text{disordered-H}$ are hydrogenated 2D sheets with various H conformations. Water molecules are present on either side of the sheet for all systems reported here. The lowest barrier for each material, is indicated in bold.}
\end{table}

To conclude, we have reported a theoretical study on proton transfer through graphene and h-BN.
After considering various factors that could impact on the penetration barriers for protons, we find that $\text{sp}^3$ hybridization at the penetration site, achieved here through hydrogenation, plays a key role
in reducing these barriers to less than 1.0 eV.
The physical origin of the barrier reduction is the elimination of the deep-lying chemisorption states
and the expansion of the honeycomb lattice at the penetration site.
Combining the major influence from hydrogenation and minor influence from NQEs, the experimentally
observed low proton transfer barrier can be rationalised.
Considering the fact that 2D materials can be functionalized with various elements other than H,
e.g. O, OH, F, Cl, this study suggests that there could be further scope for more controllable ion
and proton sieving.
We hope our study can stimulate further theoretical and experimental investigations in
this direction.
%

%%\textcolor{red}{is there any other implications beyond explaining the experiments}

\begin{acknowledgements}
Y.X.F., X.Z.L. and E.W. are supported by the National Basic Research Programs of China
under Grand Nos. 2016YFA0300900, 2013CB934600, the National Science Foundation of China under Grant Nos
11275008, 1142243, 11274012, 91021007, 11604092 and 11634001.
J.C. and A.M. are supported by the European Research Council under the European Union's Seventh Framework Programme (FP/2007-2013)/ERC Grant Agreement number 616121 (HeteroIce project).
A.M. is also supported by the Royal Society through a Royal Society Wolfson Research Merit Award.
The computational resources were provided by the supercomputer TianHe-1A in Tianjin, China and ARCHER from the UKCP consortium (EP/F036884/1).
\end{acknowledgements}

\end{document}

\title{Supplementary Information: Hydrogenation Facilitates Proton Transfer Through Two-Dimensional Honeycomb Crystals}
\author{Yexin Feng}
\thanks{These two authors contributed equally}
\affiliation{School of Physics and Electronics, Hunan University, Changsha 410082, P. R. China}
\author{Ji Chen}
\thanks{These two authors contributed equally}
\affiliation{Thomas Young Centre, London Centre for Nanotechnology, and Department of Physics and Astronomy, University
College London, London WC1E 6BT, United Kingdom}
\author{Wei Fang}
\affiliation{Thomas Young Centre, London Centre for Nanotechnology, and Department of Chemistry, University
College London, London WC1E 6BT, United Kingdom}
\author{Enge Wang}
\affiliation{International Center for Quantum Materials, School of Physics, and Collaborative Innovation
Center of Quantum Matter,Peking University, Beijing 100871, P. R. China}
\author{Angelos Michaelides}
\email{angelos.michaelides@ucl.ac.uk}
\affiliation{Thomas Young Centre, London Centre for Nanotechnology, and Department of Physics and Astronomy, University
College London, London WC1E 6BT, United Kingdom}
\author{Xin-Zheng~Li}
\email{xzli@pku.edu.cn}
\affiliation{International Center for Quantum Materials, School of Physics, and Collaborative Innovation
Center of Quantum Matter,Peking University, Beijing 100871, P. R. China}
\date{\today}

\maketitle

\section{S.I Computational details about DFT calculations}

Spin-polarized density-functional theory (DFT) calculations were performed by using the Vienna ab initio Simulation Package (VASP)~\cite{vasp1996}.
The projector-augmented plane wave (PAW) method was employed with a cut-off energy of 500 eV.
The Monkhorst-Pack (MP) k-point meshes of 2$\times$2$\times$1 and 1$\times$1$\times$1 are used for supercells with the surface periodicity of 4$\times$4 and 8$\times$8.
A slab layer of 15~\AA~thick was enough to avoid interactions between the layers.
%Since the optB88-vdW functional is used in our simulations, here we check the influence of using other functionals.
%
We performed climbing image nudged elastic band (cNEB) claculations to get the proton penetration barriers across 2D materials~\cite{cNEB2000}, with a force convergence criterion of 0.03 eV $\text{\AA}^{-1}$.
The optB88-vdW functional was chosen for the electronic structure calculations so as to obtain a good description of the dispersion forces~\cite{Klimes2010,Klimes2011}.
%
%In Fig.~\ref{figure_s1}, we present results with several other functionals, where very similar barriers are seen.
%
Here we check the influence of using other functionals on the calculated penetration barriers for proton.
We consider two kinds of penetration processes for the proton, a high barrier process with pristine graphene and a low barrier process with hydrogenated graphene, as shown in Fig. S1.
The cNEB barriers for proton penetration obtained with optB88-vdW~\cite{Klimes2010,Klimes2011}, LDA~\cite{lda1,lda2}, PBE~\cite{pbe}, PBEsol~\cite{pbe-sol}, and PBE0~\cite{pbe0-1,pbe0-2} functionals ar summarized in Table SI.
Similar cNEB barriers for proton penetration are obtained with various functionals.

\section{S.II Hydrogenated graphene and h-BN sheets}

\begin{figure}[h]
\includegraphics[width=0.65\linewidth]{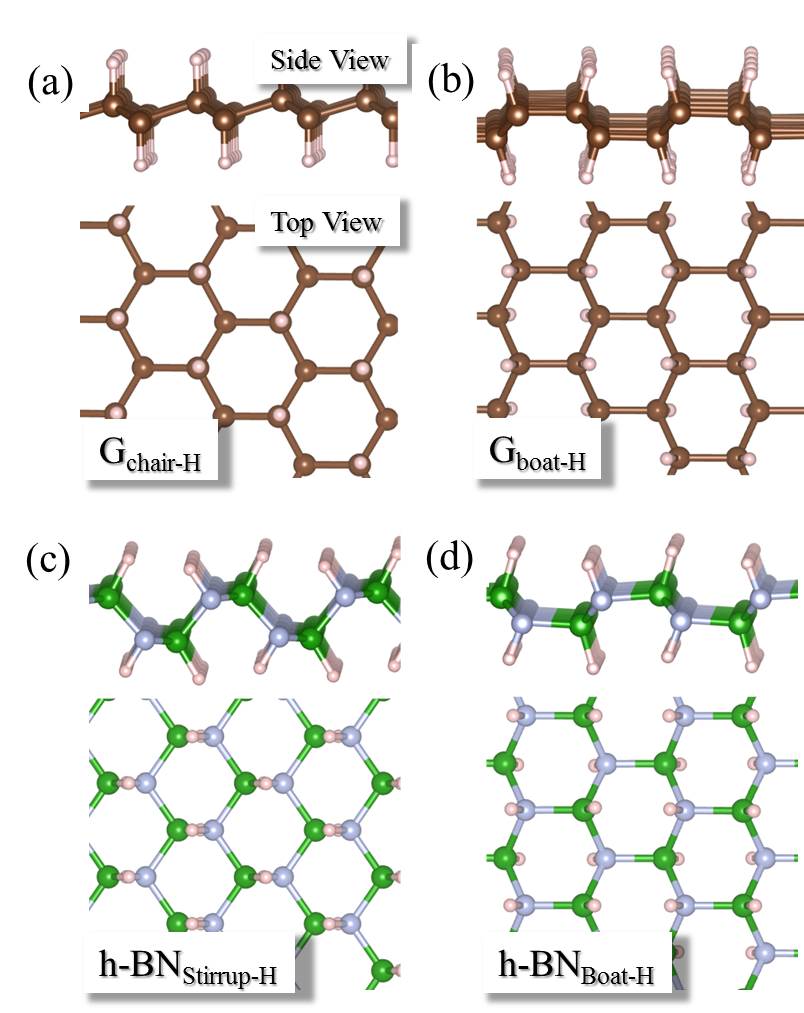}
\caption{\label{figure_s2} Atomic structures of hydrogenated graphene with (a) the chair and (b) the boat conformations and hydrogenated h-BN with (c) the stirrup and (d) the boat conformations. Brown (pink, green, gray) balls are C (H, B, N) atoms~\cite{H_cluster_prb, Samarakoon2012}.}
\end{figure}

\begin{figure}[h]
\includegraphics[width=0.55\linewidth]{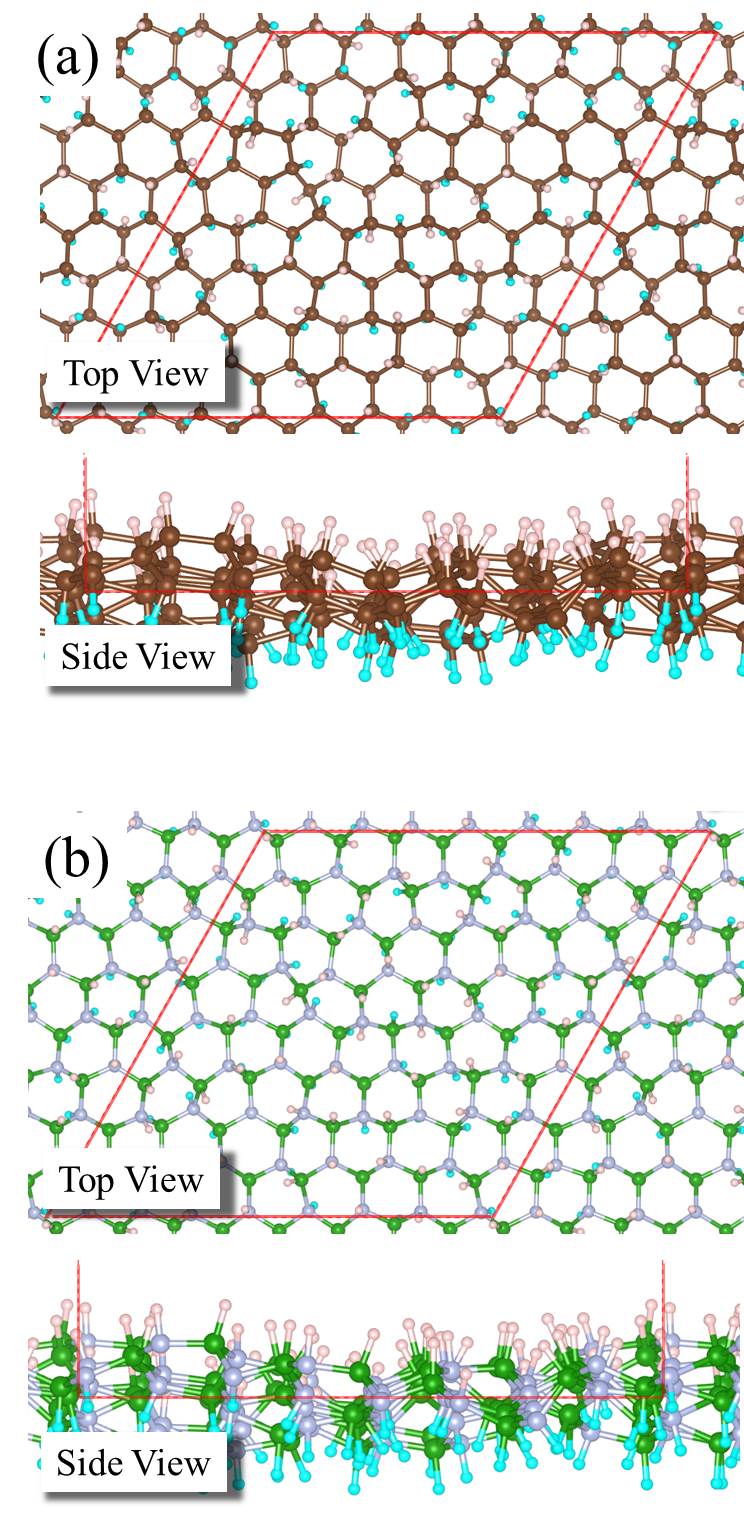}
\caption{\label{figure_s3} Atomic structures of hydrogenated (a) graphene and (b) h-BN with disordered H configurations. Brown (pink/cyan, green, gray) balls are C (H, B, N) atoms. For easy visualization, the H adatoms below the sheets are colored in cyan.}
\end{figure}

\begin{figure}[h]
\includegraphics[width=0.75\linewidth]{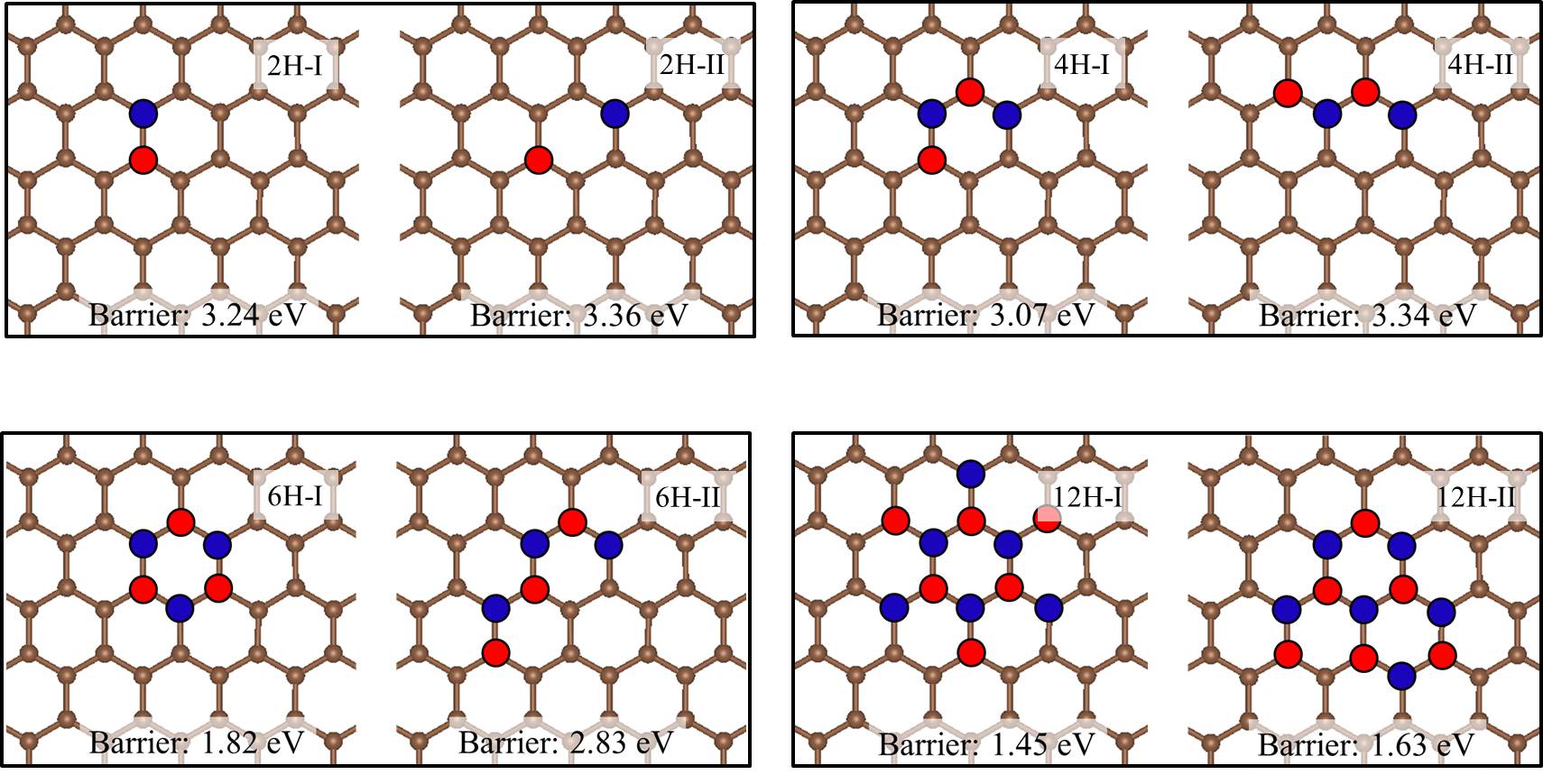}
\caption{\label{figure_s4}  The configurations of partially hydrogenated graphene with adsorbed H clusters on both sides of graphene. Red and blue dots indicate C atoms decorated by H atoms from the top and bottom side of graphene. The value of the cNEB barrier is also reported for each system. }
\end{figure}

\begin{figure}[h]
\includegraphics[width=0.95\linewidth]{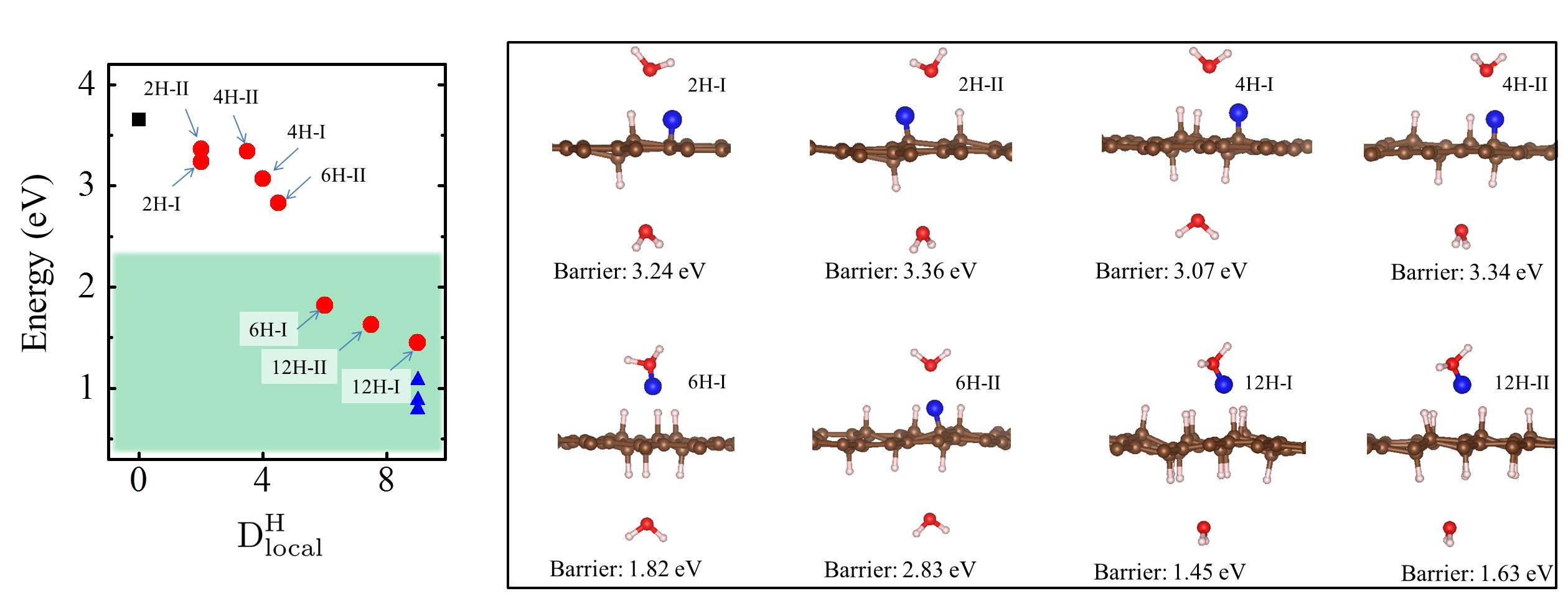}
\caption{\label{figure_s5-b} cNEB barriers as a function of the local hydrogenation degree ($\text{D}^\text{H}_\text{local}$) are shown in the left panel. The green shaded area indicates low penetration barriers with high $\text{D}^\text{H}_\text{local}$. In the right panel are the atomic structures of the initial states for the partially hydrogenated graphene as shown in the FIG~\ref{figure_s4}. Red (pink, brown) balls are O (H, C) atoms. Protons are represented by blue balls.}
\end{figure}

In the main manuscript, we have discussed the influence of hydrogenation on the proton penetration through graphene and h-BN sheets.
The atomic structures of fully hydrogenated graphene and h-BN sheets with ordered H-configurations are shown in FIG.~\ref{figure_s2}.
Apart from the ideally hydrogenated graphene sheets, we have
also studied the disordered H configurations, as shown in FIG.~\ref{figure_s3}.
A larger (8$\times$8) supercell was employed, which contains 256 atoms.
All C, N and B atoms are decorated by H atoms.
%Whether the H adatom is chemisorbed on top or below the sheet is randomly determined.
The hydrogenation sites were determined randomly on either side of the 2D membrane, and structures were fully optimized.
Besides these, the proton transfer barriers were also calculated for various partially hydrogenated graphene sheets.
For each kind of partially hydrogenated graphene, the two lowest energy structures are considered, as shown in FIGs.~\ref{figure_s4} and~\ref{figure_s5-b}.
%The atomic structures of them with calculated cNEB barriers for proton transfer are shown in FIG. S4.
%

\begin{figure}[h]
\includegraphics[width=0.75\linewidth]{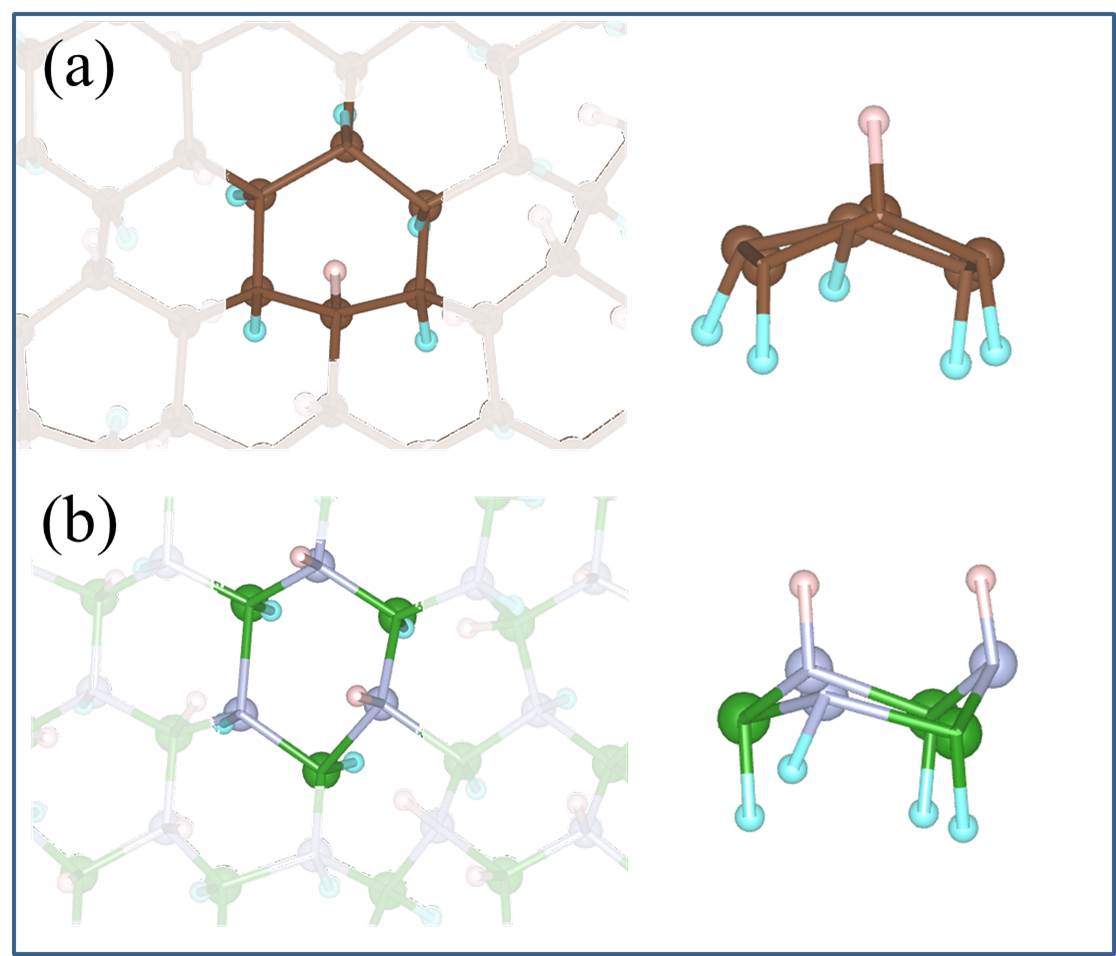}
\caption{\label{figure_s5}  The zoom-in view of atomic structures for asymmetric H decoration on hydrogenated (a) graphene and (b) h-BN with disordered H.
Red (pink, brown, green, gray) balls are O (H, C, B, N) atoms. The H adatoms below the sheets are colored with cyan (a contrast to pink) for clarity. The proton penetrates through the 2D material from the above to below of the sheet. The hexagon through which the proton penetrates is highlighted.}
\end{figure}

%
As mentioned in the main text, the asymmetric decoration of H on the opposite sides of 2D layers can further slightly reduce the proton transfer barrier.
The zoom-in view of atomic structures for this asymmetric H decoration on hydrogenated graphene and h-BN with disordered H conformations can be found in FIG.~\ref{figure_s5}.

\section{S.III PIMD SIMULATIONS}

\begin{figure}[h]
\includegraphics[width=0.85\linewidth]{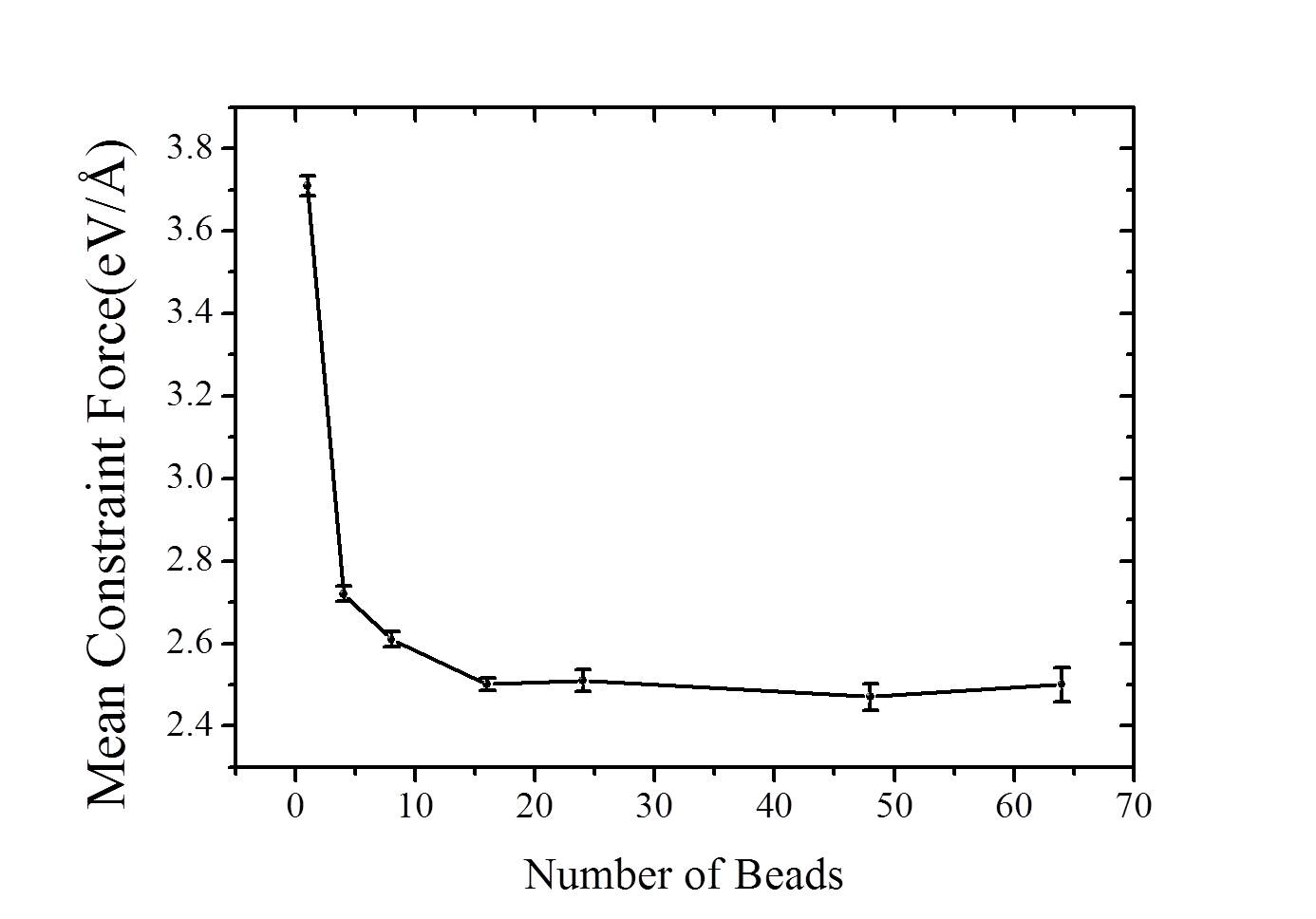}
\caption{\label{figure_s6} The mean constraint force as a function of PIMD beads, for the proton above the graphene layer with a vertical displacement of 0.18 {\AA}. A vertical displacement of 0 means the plane of the graphene layer (the C atoms fixed during the PIMD simulations).}
\end{figure}

\begin{figure}[h]
\includegraphics[width=0.75\linewidth]{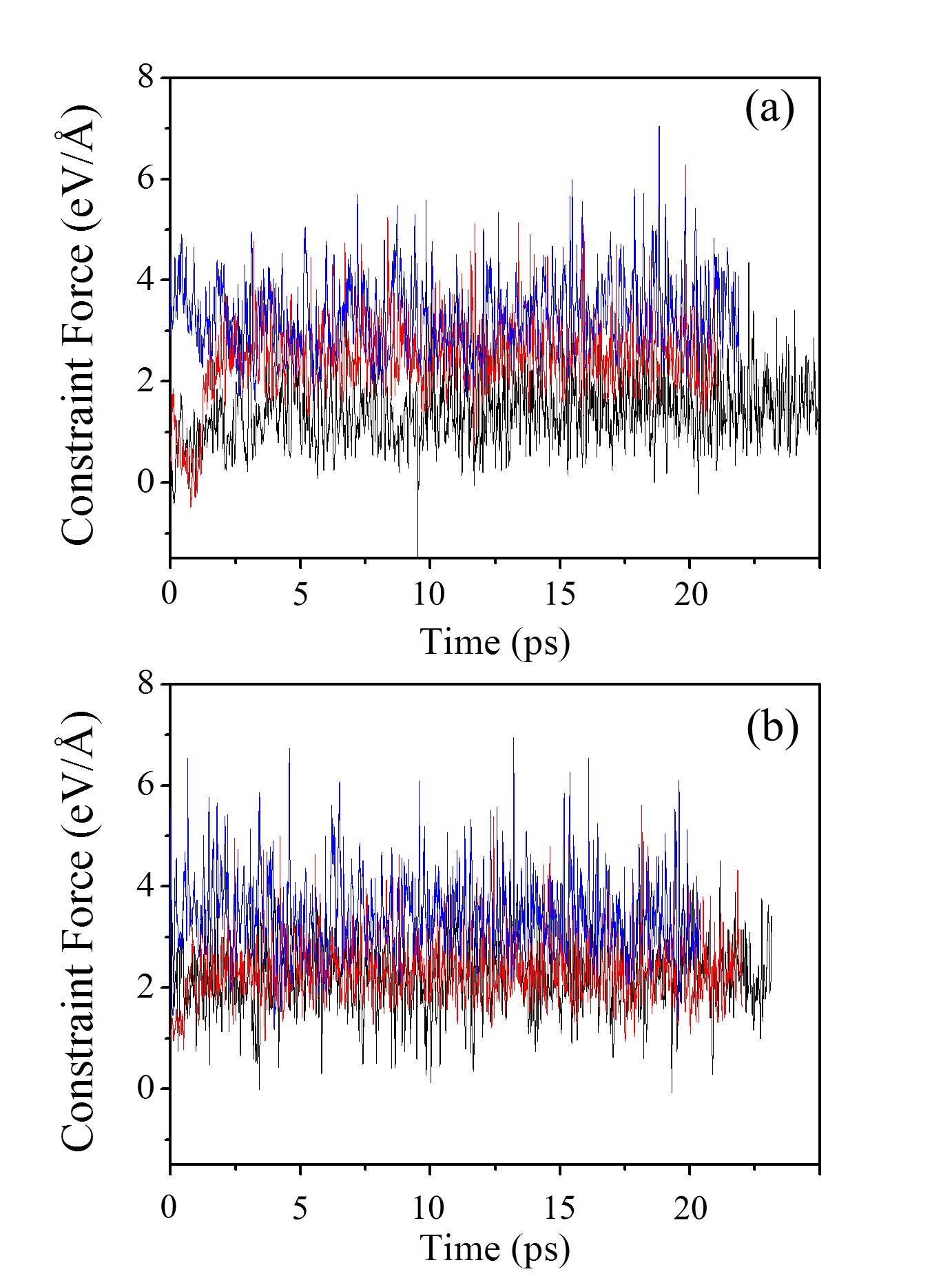}
\caption{\label{figure_s7} The constraint forces during PIMD simulations, along with the simulation time for (a) H and (b) D nuclei, at the vertical displacement of 0.0389 {\AA} (black lines), 0.1671 {\AA} (red lines) and 0.2953 {\AA} (blue lines), respectively. After thermalization ($\sim5$ ps), 30,000 steps (15 ps) were collected to calculate the constraint force, at each constraint point.}
\end{figure}

In this study, we employ constrained PIMD simulations to obtain the free energy profiles for proton penetration across graphene and h-BN sheets, by constraining the vertical displacement of proton from the 2D layers.
During the PIMD simulations, the centroids of six atoms in the hexagon through which proton penetrates are allowed to relaxed; whereas the centroids of other atoms in the 2D sheets are fixed to the positions in the pristine 2D layers.
In PIMD simulations, the number of beads used to sample the imaginary-time path-integral is a very important parameter
in descriptions of the NQEs.
We have therefore calculated the mean constraint force, for the proton above the graphene layer with the vertical displacement of 0.18 \AA, by using 1, 4, 8, 16, 24, 48 and 64 beads.
The Anderson thermostat was used to target the temperature at 300 K.
The convergence tests are shown in FIGs.~\ref{figure_s6} and~\ref{figure_s7}.
The differences between the mean constraint forces with 24, 48 and 64 beads are less
than 50 meV.
48 beads were used to for the reported PIMD results.
In the main manuscript we have shown the impact of NQEs on the free energy profile for proton transfer across pristine graphene.
NQEs reduce the transfer barrier by 0.46 eV.
In FIG. S9, we also plot the classical and quantum free energy profiles for proton penetration through the h-BN sheet.
The reduction of barrier due to NQEs is 0.28 eV.

\section{S.IV Isotope effect and Zero point energy (ZPE) correction}

\begin{figure}[h]
\includegraphics[width=0.8\linewidth]{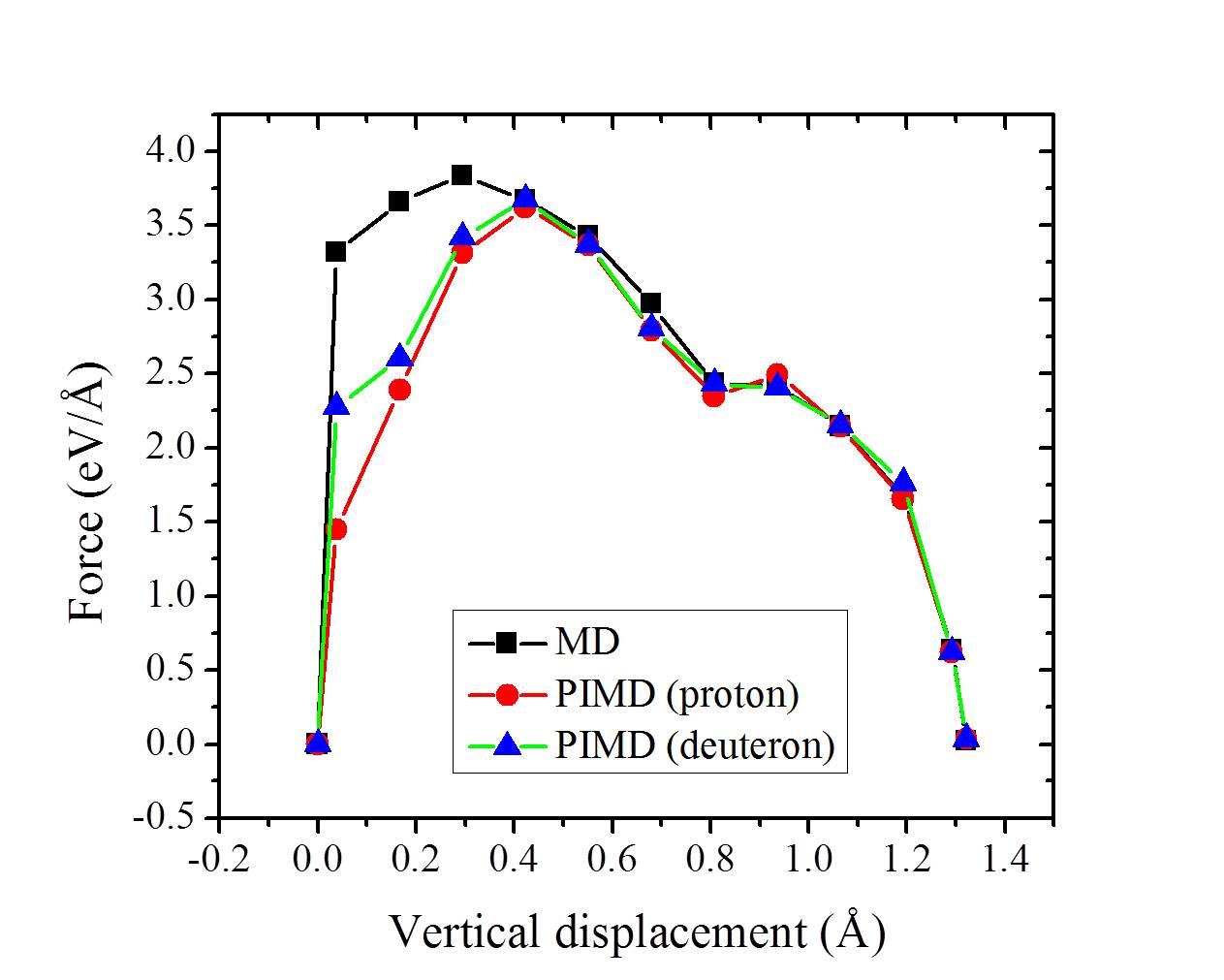}
\caption{\label{figure_s9} The classical and quantum (proton/deuteron) constraint force profiles obtained with \textit{ab initio} constrained MD and PIMD simulations. The standard errors of these mean forces are smaller than 90 meV/\AA.}
\end{figure}

\begin{figure}[h]
\includegraphics[width=0.75\linewidth]{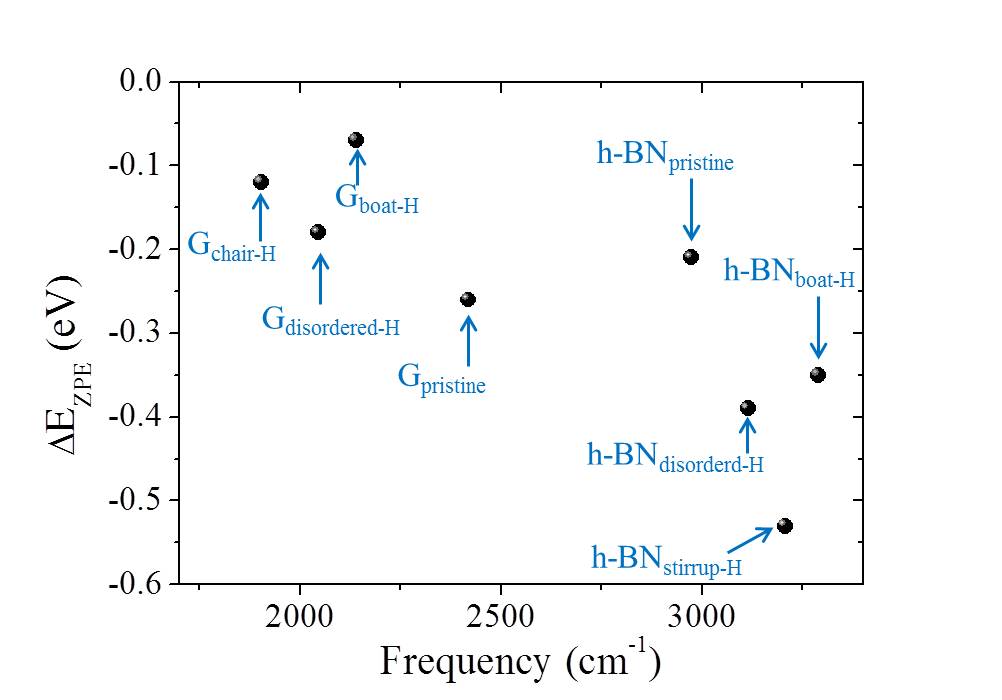}
\caption{\label{figure_s13} The calculated ZPE correction (${\bigtriangleup}E_{\text{ZPE}}$) along with the stretching frequencies of the covalent bonds between proton and O, C and N atoms in the initial state of each system studied (pristine and hydrogenated 2D sheets).}
\end{figure}

\begin{figure}[h]
\includegraphics[width=0.75\linewidth]{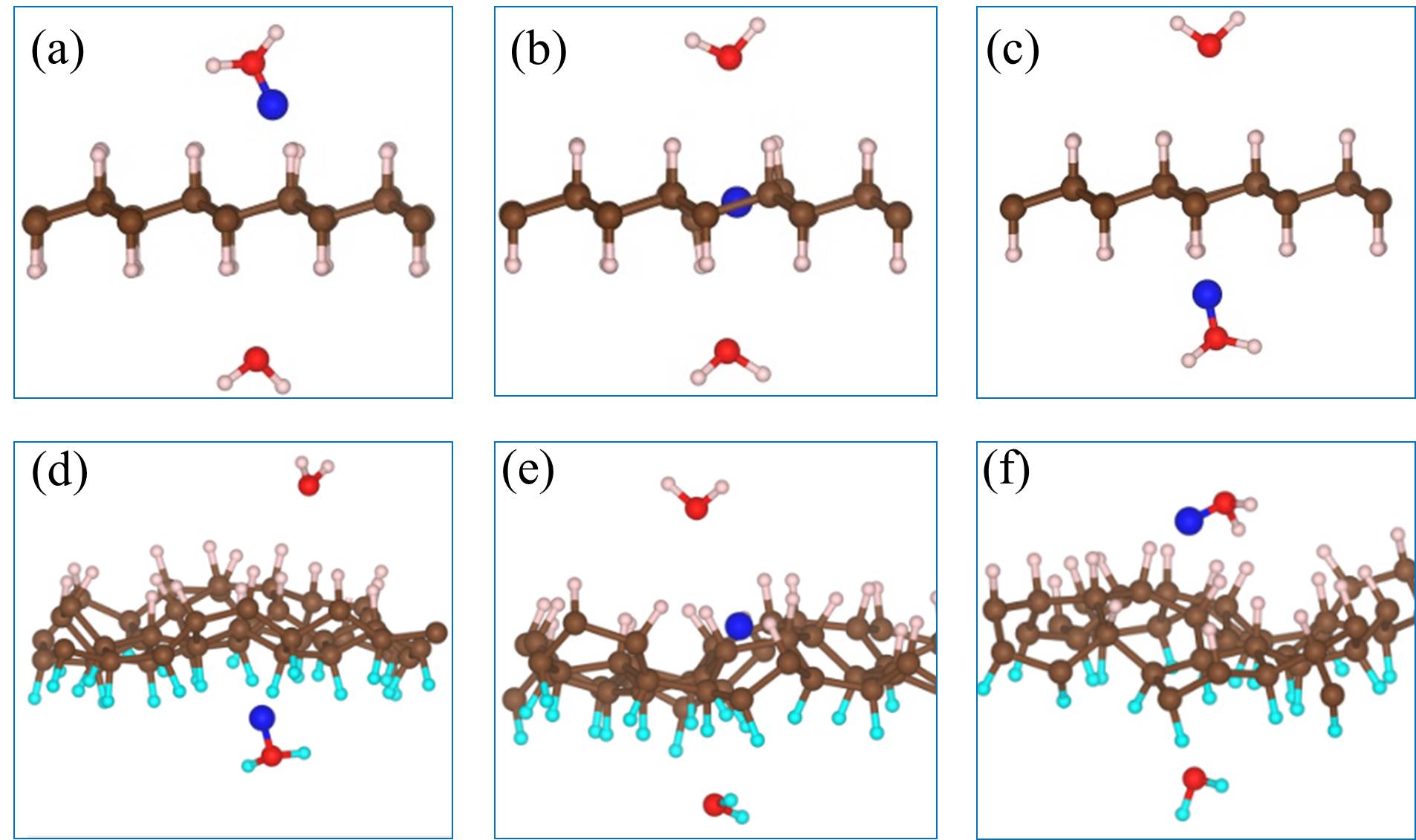}
\caption{\label{figure_s14} The close-up views of the insets in the Fig. 4 of the main manuscript, which are the atomic structures for the initial, transition and final states for fully hydrogenated graphene (a)-(c) with the chair conformation and (d)-(f) with the disordered conformation. Red (pink, brown) balls are O (H, C) atoms. Protons are represented by blue balls. The H adatoms below the sheets are colored with cyan (a contrast to pink) for clarity.
}
\end{figure}

\begin{figure}[h]
\includegraphics[width=0.9\linewidth]{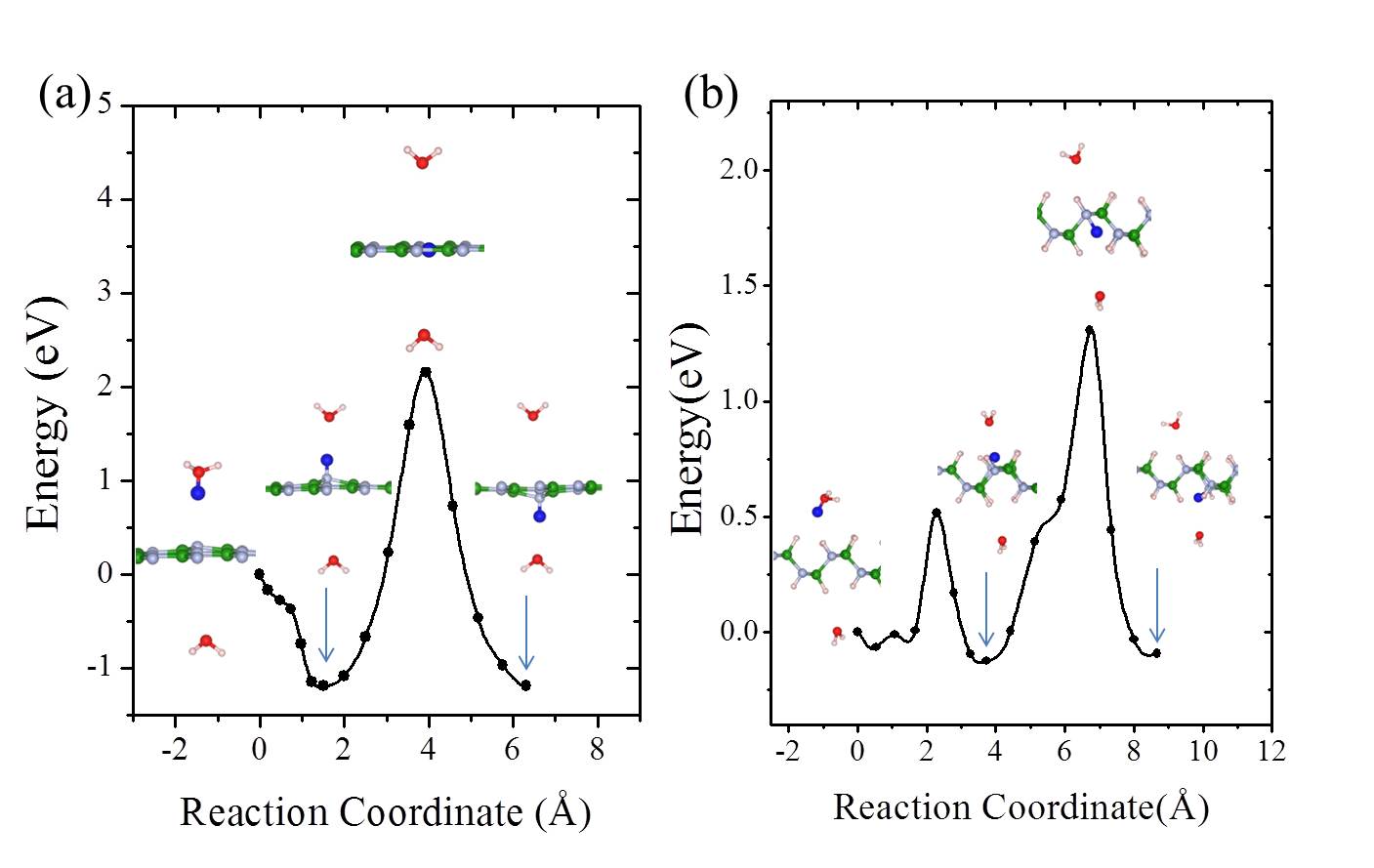}
\caption{\label{figure_s15} Energy profiles for proton penetration across (a) pristine h-BN and (b) fully hydrogenated h-BN with the stirrup conformation, in the presence of water molecules. Insets show the atomic structures for some of the key states involved in the proton penetration process. Red (pink, green, gray) balls are O (H, B, N) atoms. Protons are represented by blue balls. Hydrogenation results in the fact that
the chemisorbed state is no longer deep-lying, as compared with the pristine h-BN. This effect holds for
all hydrogenated structures discussed in this manuscript.
}
\end{figure}

\begin{figure}[h]
\includegraphics[width=0.85\linewidth]{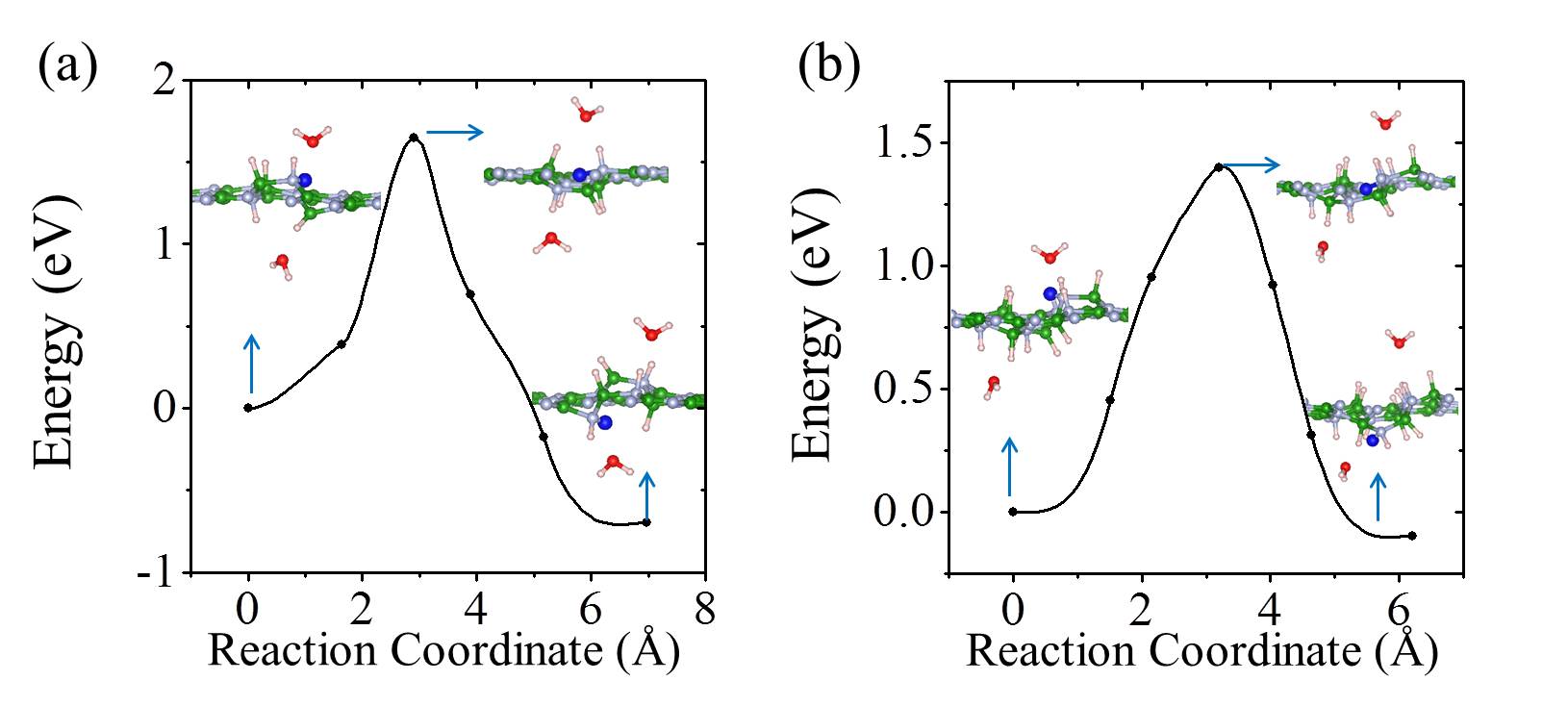}
\caption{\label{figure_s16} Energy profiles for proton penetration across partially hydrogenated h-BN with $\text{D}^\text{H}_\text{local}$ equal to (a) 6 and (b) 9, respectively, when the weighting factor w is set to 0.5. Insets show the atomic structures for the initial, transition and final states. Red (pink, brown, green, gray) balls are O (H, C, B, N) atoms. Protons are represented by blue balls.
}
\end{figure}

\begin{figure}[h]
\includegraphics[width=0.75\linewidth]{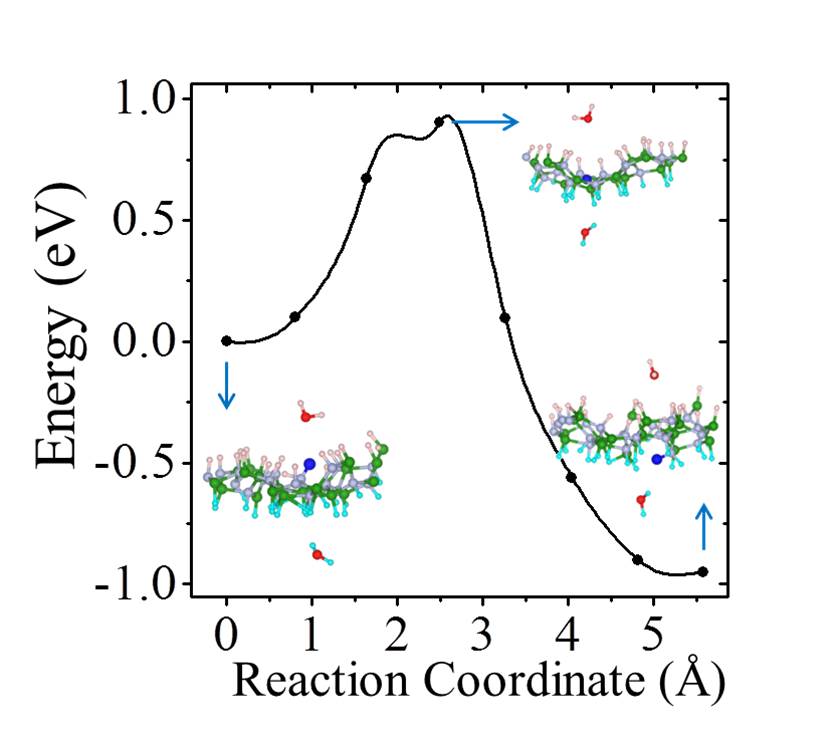}
\caption{\label{figure_s17} Energy profiles for proton penetration across fully hydrogenated h-BN with a disordered H configuration. Insets show the atomic structures for the initial, transition and final states. Red (pink, brown, green, gray) balls are O (H, C, B, N) atoms. Protons are represented by blue balls. The H adatoms below the sheets are colored with cyan (a contrast to pink) for clarity.
}
\end{figure}

The profiles of the mean constraint forces, for the proton and deuteron to penetrate through pristine graphene, are illustrated in FIG.~\ref{figure_s9}.
By integrating over the constraint forces, the free energy profiles are obtained for H and D, respectively.
For this specific process, the isotope effect is about 134 meV.
In Table I of the manuscript, for various proton penetration processes, we investigate the impact of
NQEs on the penetration barrier by doing ZPE corrections for the initial and final states.
%
%Based on cNEB calculation results for proton penetration, we further included the impact of NQEs by doing ZPE corrections.
%
The zone centered (Gamma-point) frequencies $\omega_i$ are calculated with the finite displacement
method.
The ZPEs of the initial and the final states are evaluated by $\sum \frac12\hbar\omega_i$ using
all frequencies.
${\bigtriangleup}E_{\text{ZPE}}$ is calculated as the ZPE difference between the initial and the
transition states.
We can find that for hydrogenated h-BN sheets ${\bigtriangleup}E_{\text{ZPE}}$s are larger.
This is related to the distinct atomic structures and vibrational frequencies of the initial states for hydrogenated h-BN sheets.
On h-BN, although the deep-lying chemisorbed state is avoided upon hydrogenation, the elevated
chemisorbed state is still a well-defined local minimum and we take it as the initial state
for the calculation of the barriers.
In Fig.~\ref{figure_s13}, the calculated ${\bigtriangleup}E_{\text{ZPE}}$s along with the stretching frequencies of $\text{O}\text{-}\text{H}^+$ bonds and $\text{C/N}\text{-}\text{H}^+$ bonds in the initial state of each system considered are shown.
One difference between the proton penetration mechanism through hydrogenated h-BN and that through hydrogenated graphene is that with fully hydrogenated graphene proton passes through the C ring
directly from $\text{H}_\text{3}\text{O}^+$, while with fully hydrogenated h-BN
the positively charged proton first adsorbs on negatively charged N before passing through the sheet, as shown in Figs.~\ref{figure_s14}-\ref{figure_s17}.
As the stretching frequencies of $\text{O}\text{-}\text{H}^+$ bonds in $\text{H}_\text{3}\text{O}^+$ are smaller than those of $\text{C}\text{-}\text{H}^+$ and $\text{N}\text{-}\text{H}^+$ bonds on 2D layers, ${\bigtriangleup}E_{\text{ZPE}}$'s are larger in hydrogenated h-BN systems, as shown in Fig.~\ref{figure_s13}.

\begin{figure}[h]
\includegraphics[width=0.75\linewidth]{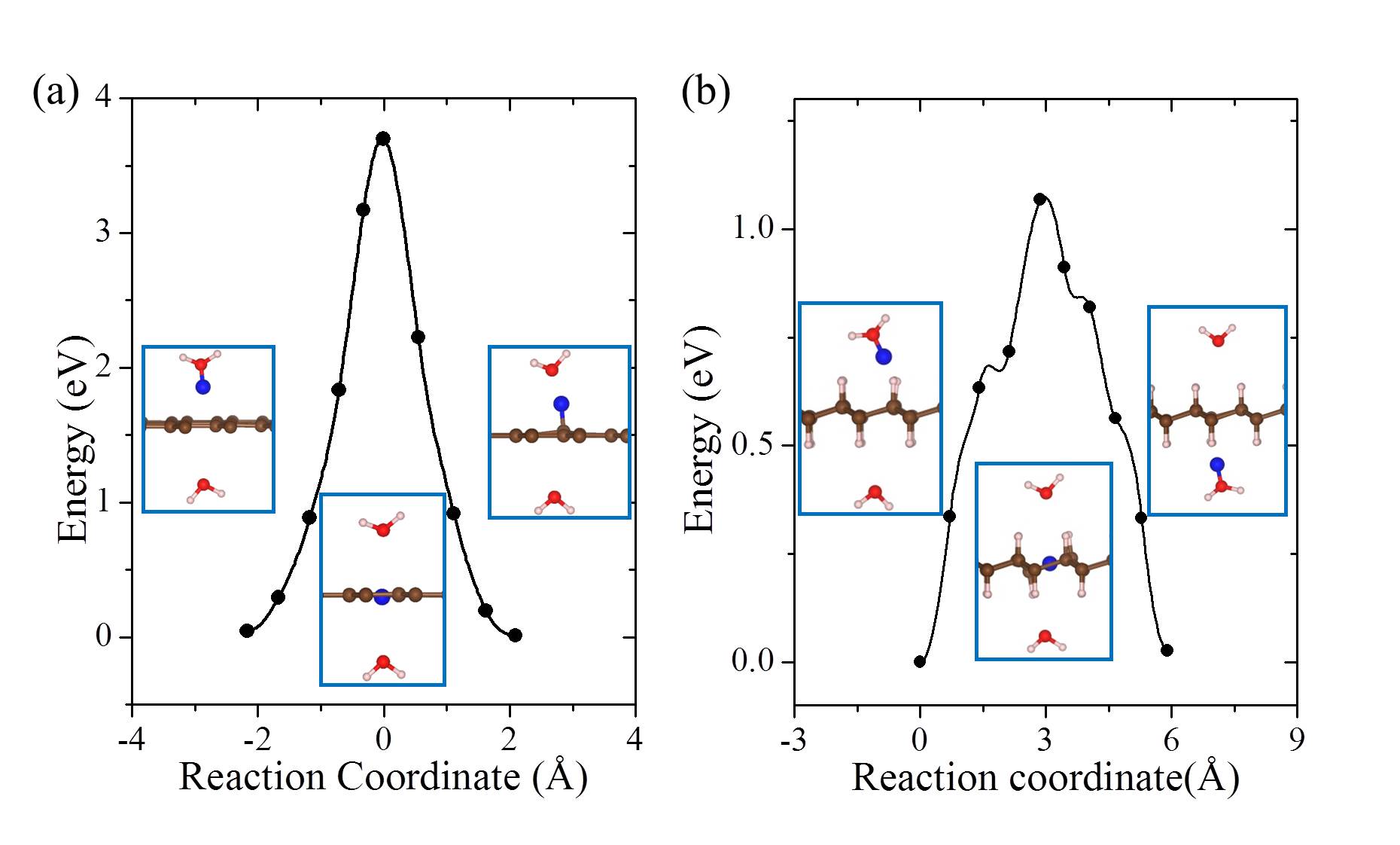}
\caption{\label{figure_s1} (a) Energy profiles for proton penetration through pristine graphene. (b) Energy profiles for proton transfer across fully hydrogenated graphene with the chair conformation. Water molecules are present on either side of the sheet for the two systems considered here.  The insets show the initial, transition and final states. Brown (red, pink) balls are C (O, H) atoms. Protons are represented by blue balls.}
\end{figure}

\begin{table}[h]
\includegraphics[width=0.5\linewidth]{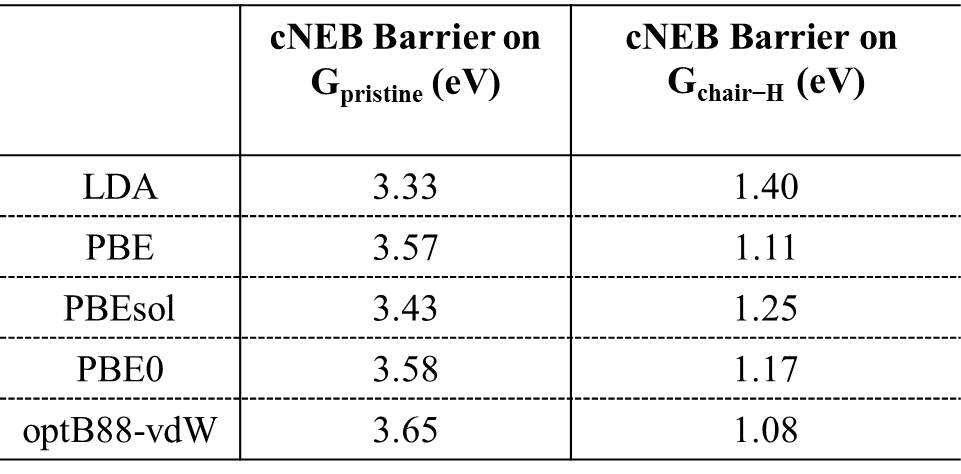}
\caption{\label{Table_s1} Calculated cNEB barriers for proton penetration through pristine graphene and fully hydrogenated graphene with the chair conformation, by using various functionals.}
\end{table}

\begin{figure}[h]
\includegraphics[width=0.75\linewidth]{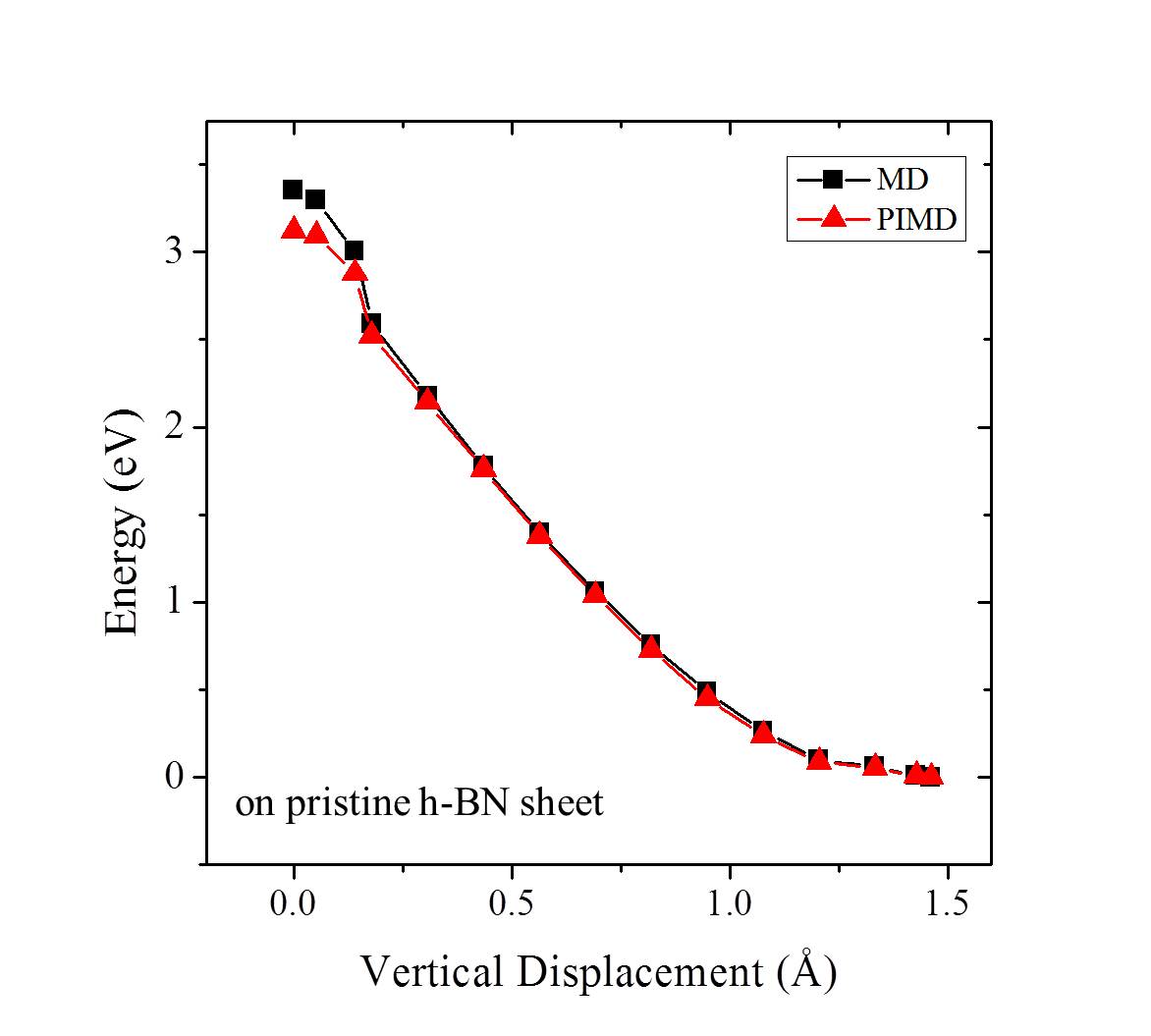}
\caption{\label{figure_s8} The classical and quantum free energy profiles for proton transfer across single-layer pristine h-BN sheet, based on constrained MD/PIMD simulations.}
\end{figure}

\begin{figure}[h]
\includegraphics[width=0.75\linewidth]{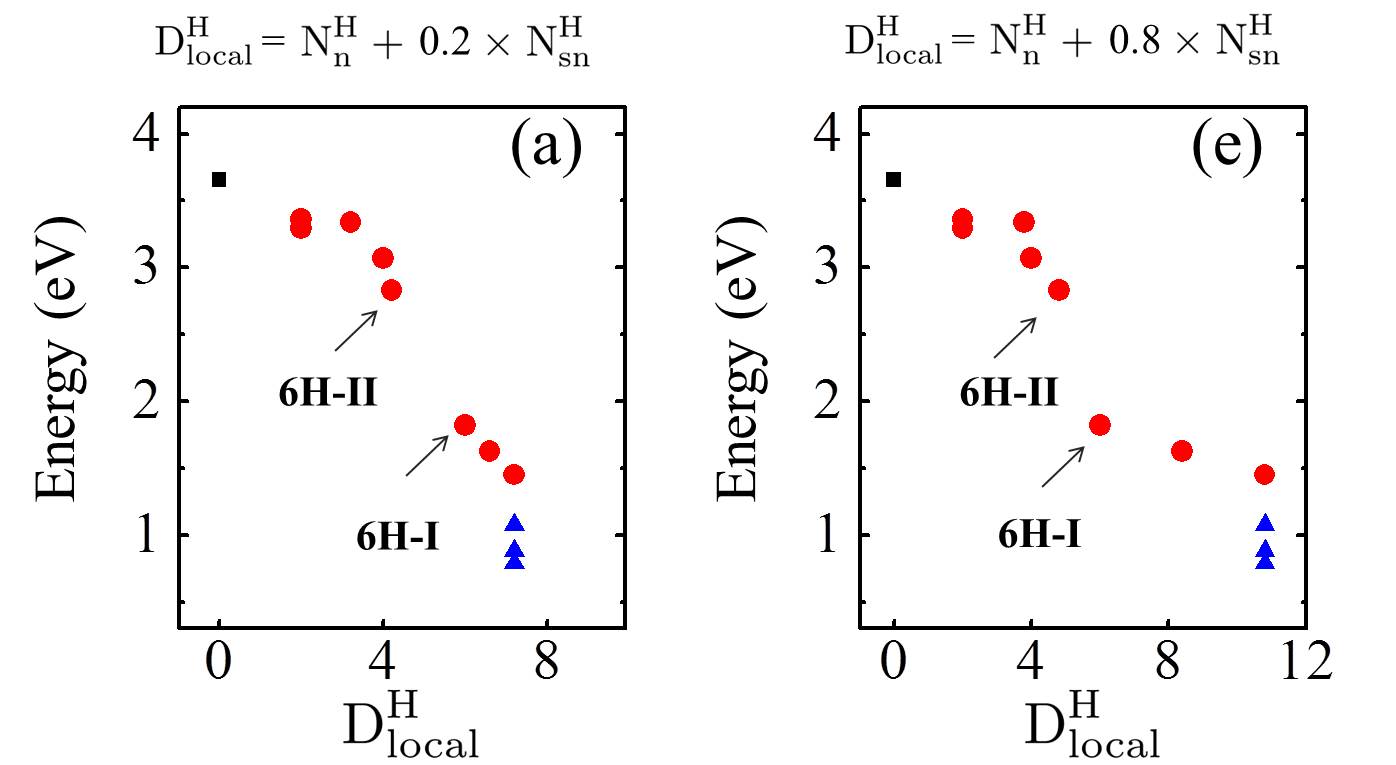}
\caption{\label{figure_s12} The cNEB barriers for partially hydrogenated graphene, as a function of the local hydrogenation
degree ($\text{D}^\text{H}_\text{local}$), with different weight factor of 0.2 and 0.8.}
\end{figure}

%\begin{figure}[h]
%\includegraphics[width=0.8\linewidth]{./fig_s10.jpg}
%\caption{\label{figure_s10} For transition states of proton transfer across 6H-I and 6H-II layers, the charge redistribution are calculated by comparing the %charge densities of hydrogenated graphene layers with and without proton. The isosurface contours of differential charge density upon the including of proton %are shown with the charge density value of $\pm$0.002 e/$\text{\AA}^3$ (yellow is density gain and cyan is density loss). The 2D contours differential charge %density at the vertical planes of proton are also shown for 6H-I and 6H-II layers in the left pannel. It is clear that the local chemical bonding at the TS of %6H-I is stronger than that of 6H-II}
%\end{figure}

%\begin{table}[h]
%\includegraphics[width=0.8\linewidth]{./Table_s1.jpg}
%\caption{\label{Table_s1} Calculated cNEB barrier, impact of NQEs on barrier (${\bigtriangleup}E_{NQEs}$) and barrier with correction ($Barrier_{correction}$) %for proton transfer across pristine and hydrogenated graphene/h-BN sheets. For pristine graphene and h-BN, ${\bigtriangleup}E_{NQEs}$ is calculated from free %energy profiles; for hydrogenated sheets, ${\bigtriangleup}E_{NQEs}$ is estimated as the zero point energy differences between initial and transition states.}
%\end{table}